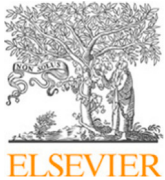
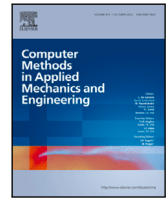

# A novel design update framework for topology optimization with quantum annealing: Application to truss and continuum structures

Naruethep Sukulthanasorn [a], Junsen Xiao [b], Koya Wagatsuma [b], Reika Nomura [a], Shuji Moriguchi [a], Kenjiro Terada [b,a],*

[a] *International Research Institute of Disaster Science, Tohoku University, Aramaki 468-1, Aza-Aoba, Aoba-ku, Sendai, 980-8572, Japan*
[b] *Department of Civil and Environmental Engineering, Tohoku University, Aramaki 6-6-06, Aza-Aoba, Aoba-ku, Sendai, 980-8579, Japan*

A B S T R A C T

This paper presents a novel design update strategy for topology optimization, as an iterative optimization. The key contribution lies in incorporating a design updater concept with quantum annealing, applicable to both truss and continuum structures. To align with density-based approaches in topology optimization, these updaters are formulated through a multiplicative relationship to represent the design material and serve as design variables. Specifically, structural analysis is conducted on a classical computer using the finite element method, while quantum annealing is utilized for topology updates. The primary objective of the framework is to minimize compliance under a volume constraint. An encoding formulation for the design variables is derived, and the penalty method along with a slack variable is employed to transform the inequality volume constraint. Subsequently, the optimization problem for determining the updater is formulated as a Quadratic Unconstrained Binary Optimization (QUBO) model. To demonstrate its performance, the developed design framework is tested on different computing platforms to perform design optimization for truss structures, as well as 2D and 3D continuum structures. Numerical results indicate that the proposed framework successfully finds optimal topologies similar to benchmark results. Furthermore, the results show the advantage of reduced time in finding an optimal design using quantum annealing compared to simulated annealing.

## 1. Introduction

Quantum computing (QC) has attracted significant attention as a solution method utilizing the capabilities of quantum mechanics, such as superposition, entanglement, and tunneling. Unlike classical computing, where information is represented as a unique bit (0 or 1), a quantum bit (qubit) can be in a superposition state of 0 and 1 with certain probabilities. This evolves the paradigm shift for computation; all candidate solutions can be searched simultaneously to find the most probable solutions. Regarding this concept, the computational speed is dramatically accelerated for solution searches, thanks to various remarkable works that have proven to be more efficient in terms of time complexity than classical computer algorithms. Notable examples include Shor's algorithm for factorization [1], Grover's algorithm for searching [2], and the Harrow–Hassidim–Lloyd (HHL) algorithm for solving linear equations [3]. Since then, several works have investigated QC algorithms to explore the used cases and the advantages of using QC to assist the solution search in various applications as a proof-of-concept stage. Currently, as the state of the art, there are two main approaches in quantum technology for implementing quantum algorithms: the quantum gate-based model and quantum annealing (QA). The quantum gate-based model is versatile, as it allows for custom operations in the algorithms through quantum circuits and logic gates, but still has a small available number of qubits, the principle is similar to how classical computers were first developed. In contrast, QA is more specific in solving combinatorial optimization problems carried out in the quantum adiabatic system, which requires a particular format, such as the Ising model or quadratic unconstrained binary optimization (QUBO), to solve the problems. Note that the chosen QC type depends on the situation at hand and the strategy for implementing quantum computation. It is worth noting that, to date, the QA-based computer has more qubits available than the gate model, allowing us to apply more practical applications. So far, QC faces several challenges due to the limitations of current quantum hardware and has not yet reached its full potential. However, quantum computer technology has made remarkable progress in recent decades, particularly with the practical application of commercial quantum computers and quantum-inspired simulators. Examples include the IBM Quantum Computer [4], D-Wave [5,6], Fujitsu Digital Annealer [7], Hitachi CMOS Annealing Machine [8], Toshiba Simulated Bifurcation Machine [9], and Fixstars Amplify Annealing Engine [10]. These advances have broadened the range of feasible applications and led to increasingly practical uses [11–19].

One of the promising applications is the development of QC-based algorithms to solve computational mechanics problems [20–22], as initially demonstrated through its successful application to solve linear equations [3]. Subsequently, several works have developed quantum algorithms for various subdisciplines within computational mechanics. In fluid dynamics, for example, QC-based methods have been proposed to solve complex fluid mechanics problems [23–26], offering new insights into turbulence and flow dynamics. Another area of application is in solving the advection-diffusion equation [27–29], which is crucial to modeling heat transfer and mass transport phenomena. Furthermore, quantum algorithms have been developed successfully for finite element (FE) analysis [30–34], a fundamental technique in computational mechanics and materials science. These works demonstrate the promising potential of quantum computing for practical use in solving complex problems as a numerical tool for quantum computer-aided engineering (Q-CAE). As quantum hardware continues to advance, these algorithms are expected to become increasingly viable for real-world engineering and scientific applications.

In addition to the aforementioned computational mechanics problems, topology optimization is one of the most attractive applications because of its robustness and versatility for various industrial sectors. The obtained results serve as a blueprint for prototyping, but efficiency and performance depend on the chosen method. As a result, various optimization algorithms have been developed, including gradient-based methods [35,36], Simulated Annealing (SA) [37,38], Genetic Algorithms [39,40], Harmony Search [41], and Evolutionary Structural Optimization [42–44]. In an effort to improve optimization techniques, QA is recognized for its potential in enhancing the solution of optimization problems. By exploiting quantum mechanical effects, the method explores solution candidates, effectively penetrating barriers in the objective function landscape. This approach demonstrates practical advantages over gate-based models, as reported in [17,45]. So far, however, there have been relatively few studies in the literature on the application of topology optimization, and two main approaches have been reported. The first approach aims to solve the overall topology optimization within a single iteration. In this approach, the optimization is typically formulated as either a ground structure or an analytical problem of the standard topology optimization. Currently, a limited number of studies have succeeded with this strategy. For example, Key and Freinberger [46] presented a formulation for optimizing the cross-sectional area of a rod under self-weight loading problems, solvable by QA. This formulation used a linear interpolation function to approximate values and cast the coefficients into QUBO format, with real-value candidate solutions encoded in binary expansion. However, their formulation was only demonstrated on a simple 1D problem with two cross-sectional area candidates. Although the first approach is promising, as it fully utilizes QA's parallel search capabilities in a single iteration, it may face hardware limitations that hinder scalability. These limitations arise from the many qubits required to encode all candidate solution patterns in the ground structure or represent real values within the formulation.

Alternatively, the second approach integrates QA within an iterative scheme, where QA functions similarly to heuristic-based optimizers in standard topology optimization, facilitating incremental updates. This iterative concept has opened a new pathway and facilitated rapid advancements in applying QA to topology optimization. For example, Wils and Chen [47,48] have explored a truss optimization problem using the D-Wave quantum annealer. Minimizing the total weight is chosen as the primary objective function, while also adhering to constraints on stress limits. In their approach, structural analysis is performed using FE analysis with a classical computer, and then QA is used to search for incremental updates to the sectional area. Their results demonstrate the feasibility of using QA to search for optimized cross-sections but are limited to nine truss members due to the limited number of available qubits. Subsequently, Wang et al. [49] expanded this framework to accommodate various boundary conditions and performed a comparative analysis between QA and SA. According to their findings, QA and SA achieved similar optimization results, but QA showed faster calculation times than SA. Recently, Honda et al. [50] presented a QA-based optimization framework for truss structures, using elastic strain energy as a Hamiltonian cost function. The cross-sectional area of the truss members was encoded into a binary expression using random number sums, and the solution was searched in an iterative manner using QA with a hybrid quantum–classical solver. This approach demonstrated the applicability of the proposed framework for large-scale problems, including 3D structures. However, while these studies have successfully applied QA to discrete truss problems, their scale and application are limited and cannot be directly extended to continuum structures.

To the best of the author's knowledge, the work by Ye et al. [51] is currently the only study that has developed a topology optimization method using QA as an iterative scheme specifically tailored for continuum structure design with high-resolution meshes. In their approach, structural analysis is initially performed using classical computing. Subsequently, general Bender decomposition (GBD) is applied to reformulate the original optimization problem into a series of mixed-integer linear programs (MILPs). A hybrid strategy combining classical computing with QA is then employed to solve optimization problems through a splitting approach. They demonstrate this method by designing the Messerschmitt–Bolkow–Blohm (MBB) beam and compare it with results from classical methods, such as the optimality criteria (OC) method. Although there is substantial evidence of QA's potential in topology optimization for continuum structures, from a practical perspective, GBD is not commonly used due to its complexities, which involve multiple stages of decomposition, cutting, and solving. In addition, the splitting approach relies on classical computers for the majority of sub-problems, reflecting a temporary strategy to leverage current quantum computing capabilities. Consequently, there remains a research gap in the literature for an efficient scheme that directly utilizes QA's results to update the design structure.

In the present study, we introduce a novel framework for design updates in topology optimization that utilizes iterative solutions from QA. This framework addresses challenges encountered in previous QA approaches, such as frameworks limited to discrete truss problems, derived analytical formulations or decomposition approaches that require solving equilibrium equations multiple times per design iteration. Specifically, the proposed framework employs update variables to directly modify the structure in each iteration and aligns with density-based approaches in topology optimization by mapping these updaters to design variables through a multiplicative formula. In addition, our proposed framework is applicable to truss and 2D–3D continuum structures, implemented on quantum annealers, simulated annealers, and hybrid solvers, aiming at its versatility and robustness for a wide range of structural optimization problems.

The organization of the paper is as follows: Section 2 provides a brief overview of the theoretical background on quantum annealing, including QA mechanics, the Ising model, annealing machines, and current progress on QA-based topology optimization. Following that, Section 3 introduces the proposed QA-based algorithm for topology optimization and presents the methodology to map the optimization framework to a QUBO model. Next, Section 4 presents numerical results to showcase the performance of the proposed topology optimization framework. Finally, Section 5 concludes the findings and identifies potential areas for further study.

## 2. Theory background

This section briefly describes the basic components of QA used to solve target optimization problems. First, the basic principle of QA adopted in this study is presented. Next, the concepts of the Ising model and binary optimization model, established to find the ground state of the solution using QA machines, are described. Subsequently, a brief overview of the available quantum and classical annealing machines to be used for the proposed quantum algorithm is provided. Finally, a review of prior research is provided, and our contribution is highlighted.

### 2.1. Quantum annealing mechanics

QA is a metaheuristic optimization technique that uses quantum mechanical phenomena to find the global minimum of a given objective function. Indeed, QA reformulates the minimization problem as the task of finding the ground state (the lowest-energy state) of a classical Ising Hamiltonian, where the Hamiltonian is a mathematical function that maps the physical system's state to its energy. The Hamiltonian encoding of the considered optimization problem is denoted as $H_{\text{problem}}$. According to the adiabatic theorem of quantum mechanics, the ground state of $H_{\text{problem}}$ can be identified by first associating it with the ground state of an initial Hamiltonian, $H_{\text{initial}}$, which serves as a starting point and is simpler to construct. By evolving the system slowly, allowing it to remain in its ground state throughout the evolution, the Hamiltonian transitions gradually from the ground state of $H_{\text{initial}}$ to the ground state of $H_{\text{problem}}$ which is the solution for considering optimization problem. The following time-dependent Hamiltonian governs the evolution:

$$H(t) = C_0(t) H_{\text{initial}} + C_1(t) H_{\text{problem}}, \qquad (1)$$

where $C_0(t)$ and $C_1(t)$ are real-valued functions that depend on the time parameter $0 \leq t \leq t_a$, where $t_a$ represents the total annealing time. At the initial time $t = 0$, the ratio of $C_1(t)$ to $C_0(t)$ is close to zero, primarily initiating the system in the ground state of $H_{\text{initial}}$. As $t$ increases, this ratio gradually shifts to guide the system toward the ground state of $H_{\text{problem}}$, with $C_1(t)$ increasing while $C_0(t)$ decreases, as illustrated in Fig. 1(a). For comprehensive details, readers may refer to Hauke et al. [52].

### 2.2. Ising model and quadratic unconstrained binary optimization (QUBO)

According to D-Wave quantum computers, the first quantum annealing machine, the Hamiltonian terms in Eq. (1) can be expressed in detail as follows:

$$H_{\text{initial}} = \sum_i \sigma_i^x, \qquad H_{\text{problem}} = -\sum_i h_i \sigma_i^z - \sum_{i<j} J_{ij} \sigma_i^z \sigma_j^z, \qquad (2)$$

where $\sigma_i^{x,z}$ are Pauli matrices operating on a qubit, $h_i$ represents the external magnetic field or qubit bias, and $J_{ij}$ denotes the interaction strength between spins $i$ and $j$ or the coupling strength. Regarding the D-wave quantum annealing process, the system begins in the ground state of the first term of Eq. (1), where all qubits are in the superposition of states 0 and 1. As the process progresses, the optimization problem is encoded in $H_{\text{problem}}$ and introduced into the system, which ideally remains in its ground state throughout the transitions (annealing) to the second term of Eq. (1), giving us the desired optimal solutions. As shown, $H_{\text{problem}}$ is the key term, tailored to the specific optimization problem being solved, namely the Ising model. Alternatively, $H_{\text{problem}}$ can be mapped into the form of QUBO through a simple transformation, as follows:

$$\sigma_i^z = 2q_i - 1. \qquad (3)$$

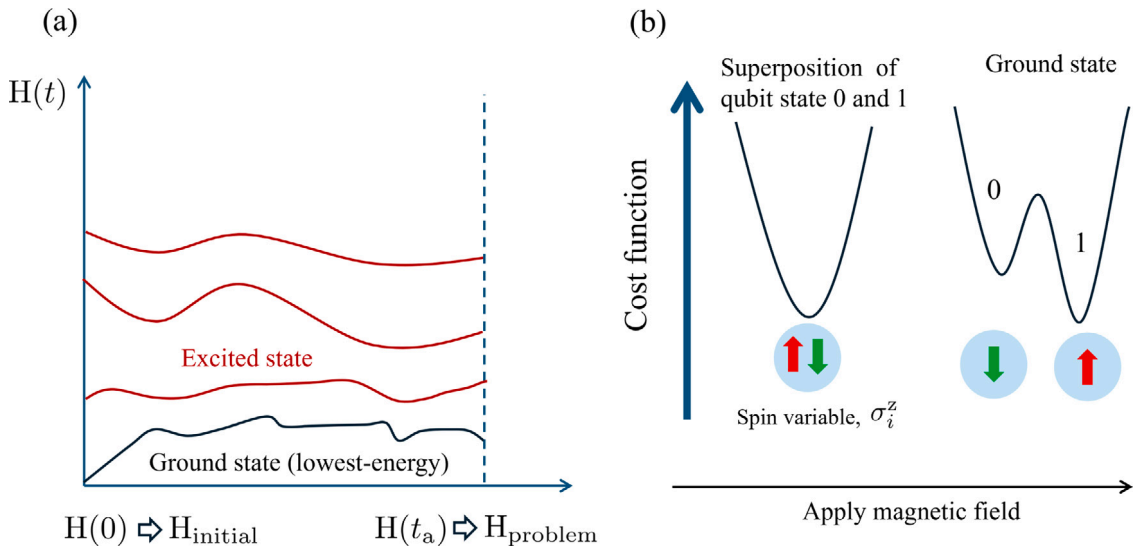

**Fig. 1.** Quantum annealing mechanics: (a) Plot of Hamiltonian energy during QA process, (b) QA operation using magnetic field within the D-Wave QPUs.

As a result, QUBO involves binary variables $q_i \in \{0, 1\}$ and can be expressed as:

$$\min_q \left( \sum_{i,j} q_i Q_{ij} q_j + \sum_i c_i q_i \right), \tag{4}$$

where $Q_{ij}$ are the elements of the quadratic term coefficient matrix, and $c_i$ are the elements of the linear term coefficient vector. Since QUBO is a convenient form for solving as it is a binary optimization problem, it is often adopted in the development of optimization algorithms using QA [53,54], which is employed in this study.

### 2.3. Annealing machines and solvers

As a frontier technology, QA machine from D-Wave company is currently the major commercially available machine for solving optimization problems using QA. D-Wave quantum computers are based on superconducting qubits, which manipulate the qubit states of 0 and 1 through circulating currents and corresponding magnetic fields, as illustrated in Fig. 1(b). Since the establishment of the first QA computer, there has been impressive progress in quantum hardware capacity over the past few decades. The quantum processing unit (QPU) has evolved from the D-Wave 2000Q, with 2000 qubits, to the new generation D-Wave Advantage system, which now boasts up to 5000 qubits. Furthermore, D-Wave has introduced a cloud computing service [55], allowing users to leverage their machines to solve problems that fit the machine's format. This service significantly accelerates the development of quantum algorithms in various applications. Despite these remarkable advancements in quantum hardware technology, there are still limitations in terms of the scalability of practical problem sizes, errors from the Noisy Intermediate Scale Quantum (NISQ) devices [56,57], and the non-deterministic nature of the solutions obtained. To address these issues, a hybrid quantum–classical solver has been implemented in D-Wave's cloud service. This solver allocates classical algorithm and the quantum computer to be capable of solving more large problems.

In parallel, specific machines inspired by the quantum annealer, known as Ising machines, have been developed to solve Ising models similar to quantum annealer. These machines can be categorized as classical annealing or quantum-inspired machines. For instance, Fujitsu's Digital Annealer [7], Hitachi's CMOS Annealing Machine [8], Toshiba's Simulated Bifurcation Machine [9], and Fixstars' Amplify Annealing Engine [10] serve as simulators that can be used to test and develop quantum algorithms. These machines provide alternative approaches to optimization by simulating quantum-inspiring annealing processes, offering valuable tools to researchers and developers in the field. The overall process using those annealing machines to solve target problems is shown in Fig. 2. Note that the optimization problem setting depends on the specific framework developed, which may address either the overall problem or a sub-problem. As can be seen, while quantum annealing machines have made significant strides, the integration of hybrid solvers and the development of Ising machines highlight the ongoing efforts to overcome current limitations and expand the capabilities of both quantum and classical optimization technologies.

In this study, the proposed design update framework is tested on these available machines not only to demonstrate the performance of the developed framework but also to prove the framework's versatility and portability as it can be run cross-platform.

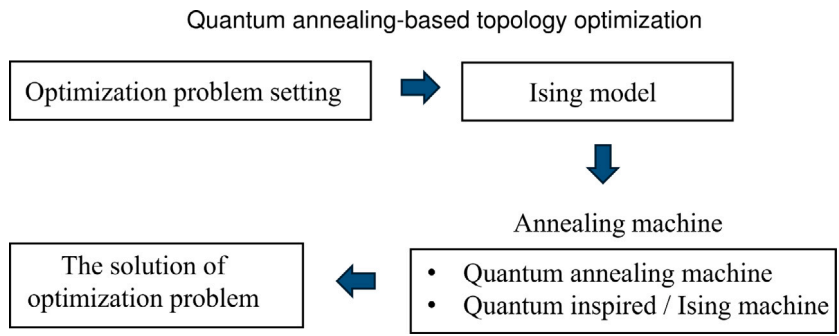

**Fig. 2.** General workflow for solving an optimization problem using annealing machines.

**Table 1**
A review of prior research on the use of quantum annealing in topology optimization.

| Name | Summary | Technique | |
|---|---|---|---|
| | | Optimization problem | Solver |
| **Analytical formulation** | | | |
| Key and Freinberger [46] | – compliance minimization<br>– 2 rod members | – 1D rod sizing optimization | D-wave: QA |
| **Iterative optimization method** | | | |
| Wils and Chen [47,48] | – weight minimization<br>– 3 to 9 truss members | – 2D truss sizing optimization | D-wave: QA |
| Wang et al. [49] | – weight minimization<br>– 5, 9, and 10 truss members | – 2D truss sizing optimization | D-wave: QA, SA |
| Honda et al. [50] | – strain energy minimization<br>– 29 and 60 truss members | – 2D and 3D truss optimization | D-wave: hybrid |
| Ye et al. [51] | – compliance minimization<br>– classical-quantum algorithm<br>– reducing problem size for QA | – 2D continuum structure<br>– GBD<br>– density-based approach | classical computer<br>D-wave: QA, hybrid |

*2.4. Current progress on QA-based topology optimization and our contributions*

In the early stages of quantum computing, several challenges remain in integrating QA into topology optimization, including issues related to scalability, noise, and non-deterministic solutions. The search for an appropriate framework that fully exploits QA's potential is still ongoing, as illustrated in Table 1. While this framework successfully showcased QA's potential for structural optimization, it was mainly focused on truss optimization. In fact, only one notable approach has utilized QA for topology optimization of continuum structures, employing the GBD method together with both classical and quantum computers. Again, GBD is less commonly used in topology optimization compared to more standard optimizers, such as the OC method [36], Method of Moving Asymptotes (MMA) [58], and Reaction–Diffusion Equations (RDE) [59], due to the complexity of solving both the master and sub-problems.

In this study, we propose a novel framework for design updates in topology optimization with the following original contribution:

- The design layout is iteratively updated by introducing the concept of a new variable, termed "updater", which is derived from solutions obtained through QA.
- The binary solution obtained from QA in the proposed framework, using only a few qubits per element, is able to effectively approximate the corresponding elemental design variable value.
- A unified design update framework using QA is proposed, designed to be applicable to large-scale problems when quantum hardware advances, while remaining flexible and applicable to truss, 2D and 3D continuum structures.

In addition, we note here that the simplicity and effectiveness of the OC method in finding optimal design have inspired the development of the current framework. However, rather than relying on optimality conditions derived from the Lagrange multiplier and gradient information to determine the updater value, we utilize QA to iteratively search for the optimal value. Furthermore, we propose a concept for developing QA-based methods in computational engineering, considering the current limitations of quantum computers, such as the number of usable qubits and noise in the solutions obtained. The development framework should include three key components: first, the developed QA-based method should be implemented on real quantum computers to validate their usability and efficiency for small-sized problems. Second, the developed method should be verified against SA or hybrid-QA approaches, which are based on similar principles but lack quantum effects, to ensure the reliability of the proposed framework. Finally, once validated with SA and hybrid-QA, the framework should demonstrate scalability by applying the proposed QA-based methods to larger or standard problem sizes using SA or hybrid-QA.

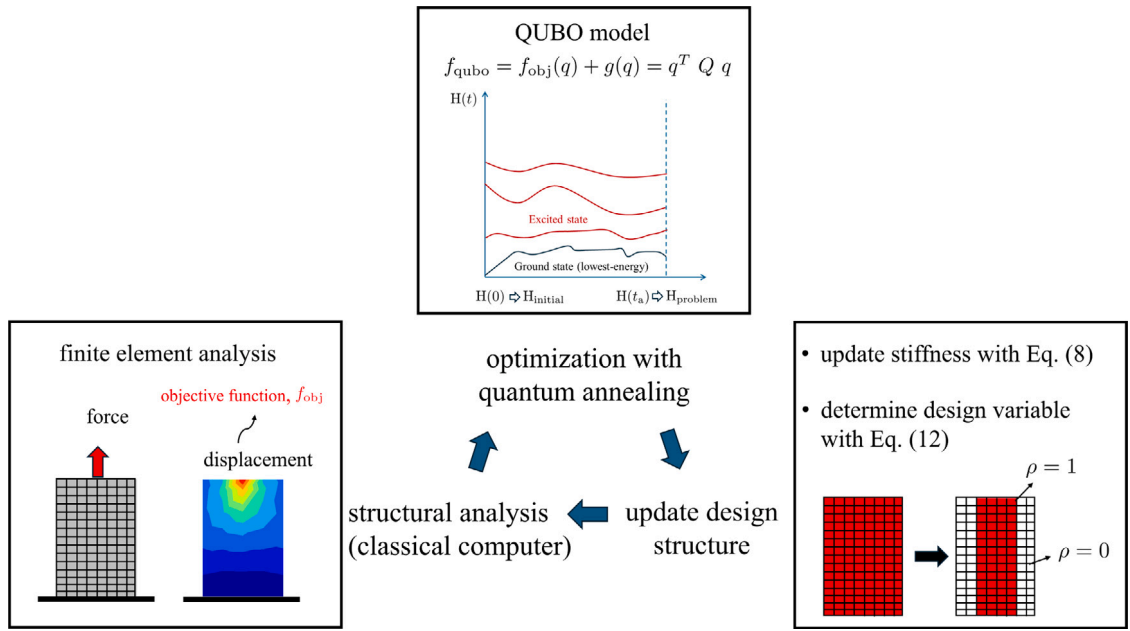

**Fig. 3.** Flow of the proposed quantum annealing framework for topology optimization.

## 3. QA-based framework for design update in topology optimization

This section presents the details of the proposed QA-based framework. First, the section outlines the overall algorithm as a series of computational steps. Next, we introduce the optimization framework for QA, which comprises an update scheme, design variables, and the setting of the optimization problem. Finally, details on the transformation of the proposed framework into a QUBO problem are presented.

### 3.1. Overall flow of proposed framework

The overall procedure of the proposed design framework is illustrated in Fig. 3 and summarized as follows:

1. Perform structural analysis using the finite element method (FEM) on a classical computer to obtain basic unknowns (e.g., displacements), and then calculate the elemental objective function.
2. Encode the elemental updaters $\alpha_e$ with binary variables $q_e$ and then determine design variables $\rho_e$. Here, subscript "$e$" indicates that the value is assigned to an element.
3. Substitute design variables $\rho_e$ into the original objective function and the volume constraint to obtain the QUBO model.
4. Solve the QUBO model for the updaters $\alpha_e$ using QA.
5. Decode the binary variables back to real values and then determine the design variable $\rho_e$.
6. Update the current design structure.
7. Repeat until convergence to an optimal design is achieved within a predetermined tolerance.

### 3.2. Update scheme and design variable

First, we introduce $\alpha$ as a variable, called an updater, for updating a design component. The value of $\alpha$ is obtained from the search of solution space through the QA process. Specifically, let $\mathbf{q} = (q_1, q_2, \ldots, q_{N_q})$ represent the set of solutions obtained from QA of $N_q$ qubits where each $q_i$ is typically a binary or spin variable. Thus, the solutions obtained via QA are mapped to $\alpha$ as follows:

$$\alpha : \mathbf{q} \mapsto \alpha(\mathbf{q}) \quad \text{with} \quad 0 \leq \alpha \leq \Theta, \tag{5}$$

where $\alpha(\mathbf{q})$ transforms the QA's binary solution $\mathbf{q}$ to a real value within the allowable range. The details of this expression will be discussed in Section 3.4. Additionally, $\Theta$ represents the maximum allowable value of $\alpha$, which is determined based on the specific problem at hand. At each $i$th design iteration, the $\alpha$ value is obtained as the solution to the optimization problem via QA and is used to update each design variable as

$$\rho^{(i)} = \alpha^{(i)} \cdot \rho^{(i-1)}, \tag{6}$$

where $\rho$ is the design variable and can be used to represent the ratio of material distribution as

$$\rho^{(i)} = \begin{cases} 1 & : \text{ solid} \\ 0 < \rho^{(i)} < 1 & : \text{ mixture} \\ 0 & : \text{ void}. \end{cases} \quad (7)$$

As in the density-based approach in topology optimization, the design variable $\rho$ can be used for the design of both truss and continuum structures through the penalty method of the general stiffness at the $i$-th design iteration as

$$K_{\text{eff}}^{(i)} = \rho^{(i)} \cdot K_{\text{o}}, \quad (8)$$

where $K_{\text{eff}}^{(i)}$ is an effective stiffness for the optimization problem of interest at $i$th iteration, and $K_{\text{o}}$ is the stiffness at the initial. Then, substitute Eq. (6) into Eq. (8) together with Eq. (7). Consequently, it can be reformulated as follows

$$K_{\text{eff}}^{(i)} = \alpha^{(i)} \cdot \rho^{(i-1)} \cdot K_{\text{o}} \quad (9)$$

$$= \alpha^{(i)} \cdot \alpha^{(i-1)} \cdot \alpha^{(i-2)} \cdots \alpha^{(1)} \cdot \rho_{\text{o}} \cdot K_{\text{o}} \quad (10)$$

$$= \prod_{j=1}^{i} \alpha^{(j)} \cdot \rho_{\text{o}} \cdot K_{\text{o}}, \quad (11)$$

$$\text{with} \quad \rho^{(i)} = \prod_{j=1}^{i} \alpha^{(j)} \cdot \rho_{\text{o}} \quad \text{and} \quad 0 \leq \prod_{j=1}^{i} \alpha^{(j)} \cdot \rho_{\text{o}} \leq 1, \quad (12)$$

where $\rho_{\text{o}}$ is the initial design variable. As can be seen, the design variable value at the $i$th iteration is the multiplicative product of updaters $\alpha^{(j)}$ from the first iteration to the current $i$th iteration and the initial design variable. In particular, at each iteration, if $\alpha = 0.5$, this iteration results in a 50% reduction in the value of $K_{\text{eff}}^{(i-1)}$. Conversely, if $\alpha = 1.1$, it means increasing the value of $K_{\text{eff}}^{(i-1)}$ by 10% as stated in Eq. (9). Through this process, the structural layout is dynamically adjusted, with unnecessary components being reduced and more critical components being reinforced. Additionally, to enforce the upper bound ($\leq 1$) stated on the right-hand side of Eq. (12), when the value of the design variable $\rho$ in the current design iteration approaches the limit value, $\Theta$ is set to 1, prohibiting $\rho$ from increasing. Fig. 4 illustrates how the real-number representation of the design variables becomes richer through the update process using $\alpha$.

### 3.3. Optimization problem setting

Thanks to the above setting, an optimization problem is established to find a design layout that achieves the target performance. In this study, the objective function is set to minimize the compliance of the structure while incorporating the material volume as the constraint. This is well known as the standard framework for structural optimization, particularly in the context of FE discretization. Once the solution is obtained, the design structure is expected to perform better than the initial design. According to this, the optimization problem can be formulated in the standard form as follows:

$$\begin{aligned} \text{find} \quad & : \quad \rho(\alpha_e) \in \{\rho_1, \rho_2, \ldots, \rho_{N_{\text{elem}}}\} \quad \text{with Eqs. (6) and (12)} \\ \min_{\rho(\alpha_e)} \quad & : \quad \mathbf{F}^\top \mathbf{U} \\ \text{s.t.} \quad & : \quad \mathbf{K}_{\text{eff}}(\rho)\mathbf{U} = \mathbf{F}, \\ & \quad \sum_{e=1}^{N_{\text{elem}}} \frac{V_e(\rho_e)}{V_0} \leq \bar{V}_{\text{target}}, \\ & \quad 0 \leq \alpha_e \leq \Theta_e, \quad 0 \leq \rho_e \leq 1, \end{aligned} \quad (13)$$

where $\mathbf{K}_{\text{eff}}$ is the global effective stiffness matrix, $\mathbf{F}$ is the external applied load vector, $\mathbf{U}$ is the global nodal displacement vector in structural analysis. Also, $\alpha_e$ and $\Theta_e$ are the elemental update updater and the corresponding maximum allowable value, respectively. It should be noted that the lower bounds of $\alpha_e$ and $\Theta_e$ are set to a small value (e.g., $10^{-6}$) to ensure numerical stability and allow for reintroduction during processing. $N_{\text{elem}}$ is the number of finite elements (or truss members), $V_e(\rho_e)$ is the elemental volume at each design iteration, $V_0$ is the initial total volume and $\bar{V}_{\text{target}}$ is a desired ratio given to the initial volume. In addition, $\rho = \{\rho_1, \ldots, \rho_{N_{\text{elem}}}\}$ is the set of elemental design variables. Here, once the optimization is established, then it will be solved for updater $\alpha_e$, via QA.

### 3.4. Mapping to QUBO model

This section provides details on formulating the optimization problem for determining the design update value in QUBO format. As a key process for solving the optimization with QA, it is necessary to derive a combinatorial optimization problem in the Ising model or, equivalently, the QUBO model. In this study, we derive an algorithm in QUBO format, which depends on each particular problem. We first redefine the cost function in Eq. (4) for the QUBO problem using the property of binary variables, such that $q_i^2 = q_i$, in the following form:

$$f_{\text{qubo}}(\mathbf{q}) = \mathbf{q}^\top \cdot \mathbf{Q} \cdot \mathbf{q}, \quad (14)$$

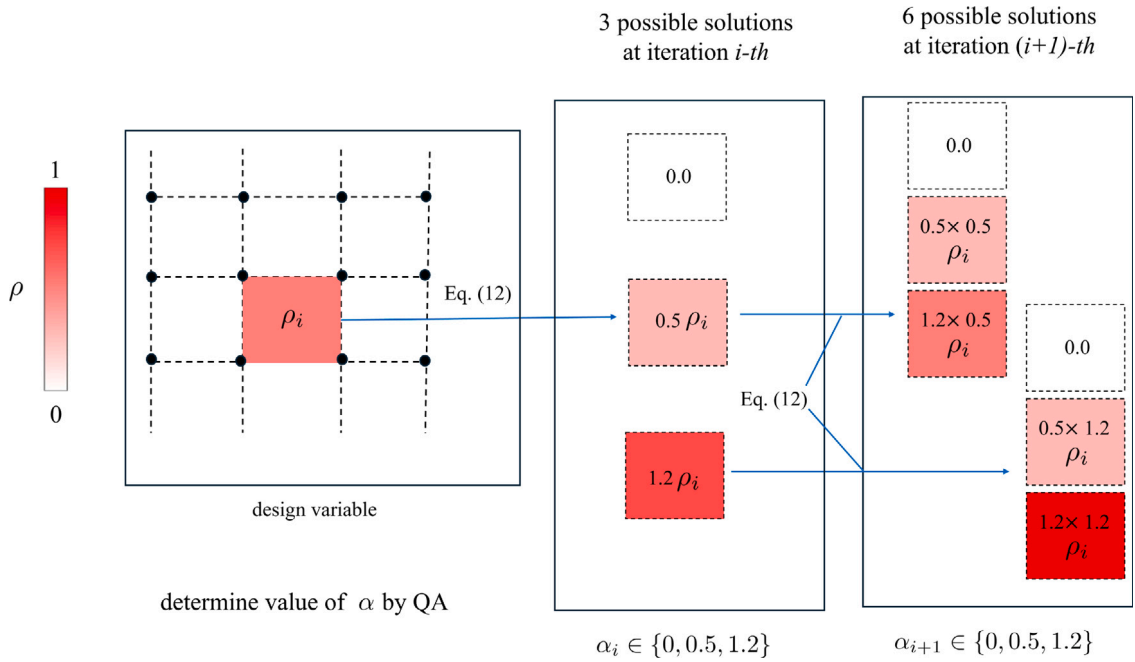

Fig. 4. Update design variable by quantum annealing.

where **Q** denotes an upper triangular coefficient matrix. Indeed, Eq. (14) can be expressed in index notation as follows:

$$f_{\text{qubo}}(q) = \sum_{k=1}^{N_q} Q_{k,k} q_k + \sum_{k<l}^{N_q} Q_{k,l} q_k q_l, \quad (15)$$

where $N_q$ is the number of qubits, $Q_{k,k}$ and $Q_{k,l}$ are the coefficients of linear and quadratic terms corresponding to the diagonal and off-diagonal entries, respectively. It can be seen that no constraint term appears in Eqs. (14) and (15), implying that the standard QUBO problem is dedicated to an unconstrained optimization problem. However, in structural optimization problems, layouts are usually designed under specific constraints to ensure optimal performance, as presented in Eq. (13). Therefore, in this study, the penalty method is employed to modify the QUBO cost function of the following form:

$$f_{\text{qubo}}(q) = f_{\text{obj}}(q) + \lambda \cdot g(q), \quad (16)$$

where $f_{\text{obj}}$ represents the objective function as defined in the structural optimization problem in Eq. (13), $g$ denotes its constraint function, and $\lambda$ is the positive penalty parameter. Note that the value of $\lambda$ must be appropriate to have an effective impact on the QUBO cost function so that the imposed constraints are satisfied.

It should be noted that since the formulation presented above has been set up as a general design framework, both the objective function and constraint can be customized and tailored to specific applications, thus achieving the desired performance. However, this study focuses on the problems of minimizing structural compliance or equivalently maximizing structural stiffness under the volume constraint. Accordingly, the objective function in Eq. (13) can be expressed using the unknown binary variable $q_e$ for each structural element as follows:

$$f_{\text{obj}}(\rho) = \underbrace{\mathbf{F}^\top \mathbf{U}}_{\text{minimizing compliance}} := \underbrace{\mathbf{U}^\top \mathbf{K}_{\text{eff}}(\rho) \, \mathbf{U}}_{\text{maximizing stiffness}}, \quad (17)$$

$$\text{with} \quad \rho = \{\rho : \alpha(q_e) \mapsto (\rho \circ \alpha)(q_e)\}. \quad (18)$$

To formulate the QUBO cost function, let us consider the $i$th iteration of the optimization process. The objective function in Eq. (17) can then be rewritten as

$$f_{\text{obj}}^{(i)}(\rho) = \left[\mathbf{U}^\top\right]^{(i)} \left[\mathbf{K}_{\text{eff}}(\rho)\right]^{(i)} \left[\mathbf{U}\right]^{(i)}. \quad (19)$$

Given that the optimization goal is to maximize overall stiffness, as defined in the objective function of Eq. (17), the QUBO cost function is formulated to identify the unknown updater $\alpha$ in the upcoming $(i+1)$-th iteration. This updater $\alpha$ is applied to adjust the stiffness, as demonstrated through Eq. (6) and Eqs. (8) to (12), guiding the design towards an optimal configuration in subsequent iterations. Consequently, by incorporating the updater into Eq. (19) together with the volume constraint, the QUBO cost function is formulated to maximize the overall structural stiffness as follows:

$$\max_{q_e} : f_{\text{qubo}}^{(i+1)} = \underbrace{[\mathbf{U}^\top]^{(i)} \boldsymbol{\alpha}(q_e)^{(i+1)} \left[\mathbf{K}_{\text{eff}}(\rho)\right]^{(i)} [\mathbf{U}]^{(i)}}_{\tilde{f}_{\text{obj}}(q_e)} + \text{volume constraint term.} \tag{20}$$

In addition, there are two key aspects to be noted for the volume constraint term. First, the considered volume constraint in Eq. (13) is an inequality constraint, so it cannot be applied in a QUBO format as it is. To address this issue, an additional positive value variable, known as the slack variable $\bar{S}$, is introduced into the inequality volume constraint in Eq. (13) as a function of another unknown binary variable $q_s$. This allows us to transform the inequality constraint to an equality constraint. For further details, refer to Glover et al. [53]. Second, to meet the requirements of the QUBO framework, $g$ is commonly formulated by squaring the volume constraint function to a quadratic form. This modification allows the volume constraint in Eq. (13) into an equality constraint at $(i + 1)$-th iteration as follows

$$\sum_{e=1}^{N_{\text{elem}}} \alpha_e^{(i+1)}(q_e) \cdot \frac{V_e^{(i)}(\rho_e)}{V_0} - \bar{V}_{\text{target}} \leq 0, \tag{21}$$

$$\sum_{e=1}^{N_{\text{elem}}} \alpha_e^{(i+1)}(q_e) \cdot \frac{V_e^{(i)}(\rho_e)}{V_0} - \bar{V}_{\text{target}} + \bar{S}^{(i+1)}(q_s) = 0, \tag{22}$$

so that the constraint function is defined as

$$g^{(i+1)}(q_e, q_s) = \left(\sum_{e=1}^{N_{\text{elem}}} \alpha_e^{(i+1)}(q_e) \cdot \frac{V_e^{(i)}(\rho_e)}{V_0} - \left(\bar{V}_{\text{target}} - \bar{S}^{(i+1)}(q_s)\right)\right)^2. \tag{23}$$

As a result, the optimization framework defined in Eq. (13) can be reformulated by adopting Eqs. (20) and (23) in the QUBO cost function to define the following minimization problem at each design iteration:

$$\text{find} : \rho \in \{\rho_1, \rho_2, \ldots, \rho_{N_{\text{elem}}}\} \quad \text{with Eqs. (6) and (12)}$$

$$\min_{q_e, q_s} : f_{\text{qubo}}(q_e, q_s) = -\mathbf{U}^\top \boldsymbol{\alpha}(q_e) \mathbf{K}_{\text{eff}}(\rho) \mathbf{U} + \lambda \cdot \left(\sum_{e=1}^{N_{\text{elem}}} \alpha_e(q_e) \frac{V_e(\rho_e)}{V_0} - (\bar{V}_{\text{target}} - \bar{S}(q_s))\right)^2$$

$$\text{subject to} \quad \mathbf{K}_{\text{eff}}(\rho)\mathbf{U} = \mathbf{F},$$

$$0 \leq \alpha_e(q_e) \leq \Theta_e, \quad 0 \leq \rho \leq 1, \quad 0 \leq \bar{S}(q_s) \leq 1. \tag{24}$$

It is worth noting that most QA computing platforms are configured for minimization-based optimization. Therefore, by using the negative of the first term in $f_{\text{qubo}}$ in Eq. (24), the problem can be reformulated such that minimizing this negative term becomes equivalent to solving a maximization problem in Eq. (20). In addition, it turns out that this expression is similar to the analytical sensitivity formulation of the density-based approach in the standard problem of minimizing compliance. Thus, from this perspective, the iterative procedure for solving Eq. (24) with QA can be considered analogous to the sensitivity analysis procedure in standard structural optimization. Note that additional constraints, such as multiple and fractional form constraints, can be incorporated into the QUBO format by introducing them as additional penalty terms. Although challenges remain to ensure effective enforcement of these constraints, addressing these issues remains a key focus for future research.

In addition, as can be seen from Eqs. (5) and (18), an explicit function is required to map the updater $\alpha$ and the design variable $\rho$ to the binary variable $q$. This process, in other words, can be recognized as an encoding process to map the real value with a particular binary function form in $n$ dimensions as

$$\alpha_e, \bar{S} : \mathbb{R} \to \{0, 1\}^n, \quad \alpha_e \mapsto q_e, \quad \bar{S} \mapsto q_s. \tag{25}$$

Particularly, the mapping function can take various forms depending on the problem formulation and the encoding scheme used in the QA process. In this study, we express it in the following form:

$$\alpha_e(q_e) = \Theta_e \cdot \xi(q_e), \quad \bar{S}(q_s) = \Theta_s \cdot \xi(q_s),$$

$$\text{with} \quad \xi(q) = \frac{\sum_{k=1}^{n_q} w_k q_k}{\sum_{k=1}^{n_q} w_k}, \tag{26}$$

where $\xi$ is a normalized binary series expansion that falls within the range [0, 1], $n_q$ represents the total number of qubits used to represent the real value function, $\Theta_s$ is a scaling factor for the slack variable, and $w_i$ denotes the predetermined weights assigned to each qubit or spin variable. In this study, uniform weights $w_k = k$ for $i = 1, 2, \ldots, n_q$ are adopted. Then, the QUBO cost function in Eq. (24) is rearranged as follows:

$$f_{\text{qubo}}^{(i+1)} = -\tilde{f}_{\text{obj}}(q_e) + \lambda \left(\sum_{e=1}^{N_{\text{elem}}} \alpha_e^{(i+1)}(q_e) \frac{V_e^{(i)}(\rho_e)}{V_0} - \left(\bar{V}_{\text{target}} - \bar{S}^{(i+1)}(q_s)\right)\right)^2$$

$$= -\tilde{f}_{\text{obj}}(q_e) + \lambda \left(\left(\sum_{e=1}^{N_{\text{elem}}} \alpha_e^{(i+1)}(q_e) \frac{V_e^{(i)}(\rho_e)}{V_0}\right)^2 - 2\left(\sum_{e=1}^{N_{\text{elem}}} \alpha_e^{(i+1)}(q_e) \frac{V_e^{(i)}(\rho_e)}{V_0}\right)\left(\bar{V}_{\text{target}} - \bar{S}^{(i+1)}(q_s)\right)\right.$$

$$+ \left(\bar{V}_{\text{target}} - \bar{S}^{(i+1)}(q_s)\right)^2\bigg)$$

$$= \underbrace{-\tilde{f}_{\text{obj}}(q_e)}_{C_1} + \lambda \Bigg( \underbrace{\sum_{e=1}^{N_{\text{elem}}} \left(\alpha_e^{(i+1)}(q_e) \frac{V_e^{(i)}(\rho_e)}{V_0}\right)^2}_{C_2} + 2\underbrace{\sum_{e<j}^{N_{\text{elem}}} \left(\alpha_e^{(i+1)}(q_e)\ \alpha_j^{(i+1)}(q_j) \frac{V_e^{(i)}(\rho_e) V_j^{(i)}(\rho_j)}{V_0^2}\right)}_{C_3}$$

$$\underbrace{- 2 \left(\sum_{e=1}^{N_{\text{elem}}} \alpha_e^{(i+1)}(q_e) \frac{V_e^{(i)}(\rho_e)}{V_0}\right) \left(\bar{V}_{\text{target}} - \bar{S}^{(i+1)}(q_s)\right)}_{C_4} + \underbrace{\left(\bar{V}_{\text{target}} - \bar{S}^{(i+1)}(q_s)\right)^2}_{C_5} \Bigg). \tag{27}$$

Then, the substitution of Eqs. (12), (24) and (26) into Eq. (27) yields

$$C_1 = -\sum_{e=1}^{N_{\text{elem}}} \Theta_e \cdot U_e^\top \cdot \xi(q_e) \cdot K_e \cdot U_e, \tag{28}$$

$$C_2 = \sum_{e=1}^{N_{\text{elem}}} (\Phi_e)^2 \cdot \xi^2(q_e), \tag{29}$$

$$C_3 = 2 \sum_{e<j}^{N_{\text{elem}}} \Phi_e \cdot \Phi_j \cdot \xi(q_e) \cdot \xi(q_j), \tag{30}$$

$$C_4 = -2 \left(\sum_{e=1}^{N_{\text{elem}}} \Phi_e \cdot \xi(q_e)\right) \bar{V}_{\text{target}} + 2 \sum_{e=1}^{N_{\text{elem}}} \left(\Phi_e \cdot \Theta_s \cdot \xi(q_e) \cdot \xi(q_s)\right), \tag{31}$$

$$C_5 = \bar{V}_{\text{target}}^2 - 2\bar{V}_{\text{target}} \cdot \Theta_s \cdot \xi(q_s) + \Theta_s^2 \cdot \xi^2(q_s), \tag{32}$$

$$\text{with} \quad \Phi_k = \Theta_k \cdot \left(\frac{V_k}{V_0}\right). \tag{33}$$

Using this QUBO format, the minimization problem in Eq. (24) can be solved on a QA computing platform.

**Remark 1.** In practice, the number of available qubits on most current quantum machines is limited. Also, as the number of qubits increases, the noise and error in the quantum machine become more significant. Accordingly, it may be preferable to use as few qubits as possible at this early stage. With this in mind, the binary expansion in Eq. (26) can be simplified by setting the number of qubits $n_q$ to one. In this case, then $\xi(q) = q$, and the updater $\alpha$ can take values of 0 or $\Theta$ at each design iteration. Consequently, this approach is similar to the iterative optimization of binary variable topology (0/1), as seen in evolutionary structural optimization [42–44]. However, it uniquely incorporates scaling with $\Theta$ in each iteration, allowing topology updates.

**Remark 2.** It is worth noting that the optimal design could not be achieved by directly using the design variable $\rho$ as the primary unknown variable solved by QA in Eq. (13), as this approach might allow only a single iteration, leading to suboptimal solutions. Additional techniques, such as analytical formulations [46], incremental updates [49,50], or GBD [51], may be required to solve the original problem effectively. On the other hand, our proposed framework introduces the concept of using qubits as the primary unknown variables, which are then mapped to each updater to determine the corresponding design variable at each design iteration. This approach not only enables iterative optimization but also accommodates the need for a large number of qubits to effectively represent the design values, as demonstrated in the numerical example.

### 3.5. Algorithm

The QA framework proposed in Section 3.1 is outlined in Algorithm 1. In-house code is developed in Python, which can be implemented via the software development kits (SDKs) of quantum cloud services such as D-Wave [5,6] and Fixstars Amplify [10]. In this study, the code is developed using the Fixstars Amplify SDK, allowing us to test our code on both the Amplify Annealing Engine (Amplify AE) [10], a GPU-based Ising machine in Japan, and the quantum annealer from D-Wave, a QPU-based quantum cloud computing [5,6] in Canada.

In addition, we define "TO-qAnneal" as the version of our algorithm executed solver from D-Wave's Advantage system 6.4 quantum annealer, "TO-sAnneal" as the version running solver on Amplify AE's simulated annealing engine (a GPU-based Ising machine), and "TO-bqmAnneal" for the hybrid binary quadratic model (BQM) solver developed by D-Wave. While these implementations share the same core algorithm, they are executed on different solvers to evaluate performance and demonstrate the reliability and versatility of our proposed QA framework across multiple quantum computing platforms. In fact, since the quantum annealer is still limited in problem size, TO-sAnneal and TO-bqmAnneal can be used to test the developed QA-based algorithm on a larger scale. Notably, both a GPU-based Ising machine and hybrid QA are established as ideal platforms specifically for proof-of-concept development and practical applications in QA.

**Algorithm 1** QA-based design update strategy
---
1: Initialize parameters: $\alpha$, $\rho$, $\lambda$, $\Theta$, $V_0$
2: **while** the optimization convergence criteria are not satisfied **do**
3:     **[Structural Analysis]**
4:     Solve equilibrium equation with a classical computer
5:     **[Optimization via QA]**
6:     Encode each updater $\alpha_e$ to binary variables $q_e \to$ Eq. (26)
7:     Formulate QUBO cost function $\to$ Eqs. (27)–(33)
8:     Use QA to find the ground state of the QUBO cost function, determining solutions of binary variables $q_e$
9:     Decode solutions of binary variables $q_e$ to each updater $\alpha_e \to$ Eq. (26)
10:    Update design variables $\rho \to$ Eqs. (6) and (12)
11:    **if** $\rho > 1$ **then**
12:       Set $\Theta = 1$ and enforce the upper bound $\rho = 1$
13:    **end if**
14:    Check if the optimization convergence criteria are satisfied
15: **end while**
16: **return** the optimized design variables
---

**Remark 3.** In Algorithm 1, Lines 1–7 are executed on a local classical computer. Subsequently, the QUBO cost function is submitted to a quantum cloud computing platform in Line 8 for searching the ground state of Eq. (27) to determine the binary solutions of updater $\alpha$.

*3.6. Benchmarking time to search for the optimal design*

Comparing the computational performance of QA-based algorithms with another classical and hybrid quantum–classical methods remains challenging due to the noise present in current quantum processors. To mitigate this, repetitive runs on quantum processors can be employed to determine the most probable solution. Probabilistic performance metrics have been proposed to evaluate performance based on target accuracies with an exact solution, as detailed in [60]. However, we may desire a deterministic design solution for topology optimization, as our focus is on the obtained topology in each design iteration. In fact, issues such as the uniqueness and availability of optimization solutions can arise in certain cases, where reference structures needed for the metric may not yet exist. Direct measurements of runtime on the central processing unit (CPU) and quantum processing unit (QPU) were used for truss optimization to compare QA and SA on the same platform provided by D-Wave, as reported in [49]. However, when running on different solver platforms, variations in device architectures can complicate comparisons due to overhead factors such as encoding time, readout time, and cloud network communication. These factors contribute to the current limitations in fully realizing the full potential of QPUs. Nevertheless, significant progress has been made in reducing overhead time in recent years, and it is expected that QPUs will accelerate their speed in the near future.

Consequently, in this work, we compare the performance by focusing on the time required for the algorithm to find design update, separate from the overhead time. We refer to this as Time-to-Find-Solution (TFS), which can be calculated as follows:

$$\text{TFS}_{\text{qa}} = t_a \cdot R \cdot I_N, \tag{34}$$

$$\text{TFS}_{\text{sa}} = t_s \cdot I_N, \tag{35}$$

where $\text{TFS}_{\text{qa}}$ and $\text{TFS}_{\text{sa}}$ represent the total time spent to find all optimal designs for topology optimization via `TO-qAnneal` and `TO-sAnneal`, respectively. The term $t_a$ refers to the annealing time for the quantum annealer, while $t_s$ denotes the time to find a design update using the simulated annealer. Also, $R$ is the number of repeated samplings for QA, and $I_N$ is the total number of design iterations.

In addition, we define $T_{\text{qpu}}$ as the total time from accessing the QPU to receiving the solution, which includes all overhead within the QPU. Moreover, $t_{\text{sa}}$ and $t_{\text{bqm}}$ represent the execution time limits for `TO-sAnneal` and `TO-bqmAnneal`, respectively. It should be noted that this is not the actual time used to find the solution, but rather the maximum time allocated for the machine to search. Thus, the total time spent by the machine searching for solutions in `TO-sAnneal` and `TO-bqmAnneal` can be expressed as $T_{\text{sa}} = t_{\text{sa}} \cdot I_N$ and $T_{\text{bqm}} = t_{\text{bqm}} \cdot I_N$, respectively. Also, the OC method is run on a local computer with an Intel(R) Xeon(R) Gold 6240R 2.40 GHz CPU.

**Remark 4.** The parameter $R$, corresponding to the solver parameter "`num_reads`" in D-Wave, controls the number of repeat samples. This helps mitigate the effects of noise and disturbances from external energy sources, as real-world quantum computations cannot operate in perfect isolation. Factors such as thermal fluctuations can cause the system to shift from the desired ground state to higher energy levels, reducing the probability of remaining in the lowest-energy state for the problems considered. By generating a sufficiently large number of samples, a distribution of potential solutions is produced, allowing the most probable solution to be identified.

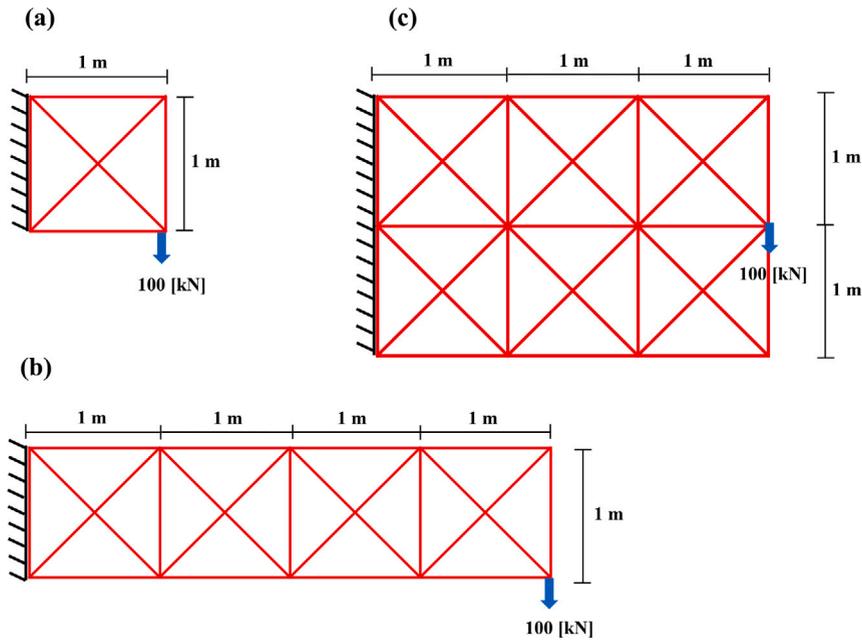

**Fig. 5.** Optimization target: three truss structures with different numbers of members.

**Remark 5.** The TFS metric may not be ideally suited for general comparing performance with the OC method due to several indirect computational costs, such as overhead, communication time, and FE analysis time which can vary depending on the total number of design iterations. Furthermore, quantum hardware is still in its early stages and has not yet surpassed classical computing in general. At this point in the ongoing development of quantum hardware, the TFS metric is primarily used to highlight the advantages of quantum effects in the solution search process, when compared to similar classes of optimization algorithms.

## 4. Numerical example

Numerical experiments are carried out to demonstrate the promise and performance of the proposed QA-based algorithm. The target designs are truss structures, and 2D and 3D continuum structures.

### 4.1. Truss optimization

We consider the optimal design problem of three truss structures with different geometries and boundary conditions, as shown in Fig. 5. As mentioned before, the state variables (e.g. displacement) are obtained by performing standard structural analysis on a classical computer. In this particular problem, the effective stiffness $K_{\text{eff}}$ stated in Eq. (8) can be a truss member $e$ as $K_e^{(i)} = \rho_e^{(i)} K_0 = \rho_e^{(i)} \cdot E \cdot A_e^0 / L_e$, where $\rho_e^{(i)}$ is the elemental design variable in the iteration $i$ of the design, representing the ratio between the current and initial cross-sectional areas, $A_e^{(i)}/A_e^0$, and $A_e^0$ is the initial cross-sectional area set at 0.5 m$^2$ for all members. Here, $E$ is the Young's modulus set at $2 \times 10^{11}$ N/m$^2$ and $L_e$ is the length set at either 1 m or $\sqrt{2}$ m as shown in the figure. The target ratio $\bar{V}_{\text{target}}$ of the total volume to the initial total volume $V_0$ is fixed at 1 throughout the optimization process, implying that $V_0$ is the volume constraint for the optimization problem. The current design volume, denoted by $V_{\text{des}}^{(i)}(\rho)$, corresponds to $\sum_{e=1}^{N_{\text{elem}}} V_e^{(i)}(\rho_e)$ where $V_e^{(i)}(\rho_e) = \rho_e^{(i)} \cdot A_e^0 \cdot L_e$. Additionally, the number of unknown binary variables or qubits can be express as:

$$N_{\text{q}} = N_{\text{elem}} \cdot n_{\text{q}} + n_{\text{s}}, \tag{36}$$

where $n_{\text{q}}$ and $n_{\text{s}}$ are the numbers of qubits for the elemental updaters $\alpha$ and the slack variable $\bar{S}$, respectively, both of which are set to 1 in this study, as mentioned in Remark 1. In the truss example, the execution timeout parameters for TO-sAnneal and TO-bqmAnneal are set as the default $t_{\text{sa}} = 10$ seconds and $t_{\text{bqm}} = 3$ seconds, respectively, and the annealing time of TO-qAnneal is $t_{\text{a}} = 20$ microseconds. Meanwhile, the maximum allowable value $\Theta$ is determined through trial and error for each specific problem, and so is the penalty constant $\lambda$. In this example, we set $\Theta_e = 1.1$ and $\Theta_s = 0.02$. The optimization iterative process ends after the objective function value changes by less than 1% for five consecutive iterations.

First, we apply TO-qAnneal, TO-sAnneal, TO-bqmAnneal, and OC method to the 6-member truss structure in Fig. 5(a). The parameters are set as $\lambda = 5$, and $\rho_0 = 0.35$. Repetition sampling is set to 200 times in each design iteration for TO-qAnneal, the limit timeout is default set at 10 and 3 s for TO-sAnneal and TO-bqmAnneal, respectively, and the number of total qubits is $N_q = 7$. The results are shown in Fig. 6. As can be seen, the objective function value from TO-qAnneal smoothly converges to the optimal design and satisfies the volume constraint within 16 iterations. In addition, the final configuration after convergence is the well-known optimum design to the two-bar truss problems [42,47]. Additionally, the identical optimized design is obtained for TO-qAnneal, TO-sAnneal, and TO-bqmAnneal, all of which implement the proposed algorithm run on different solvers. Also, the optimized topology obtained agrees well with that of the OC method.

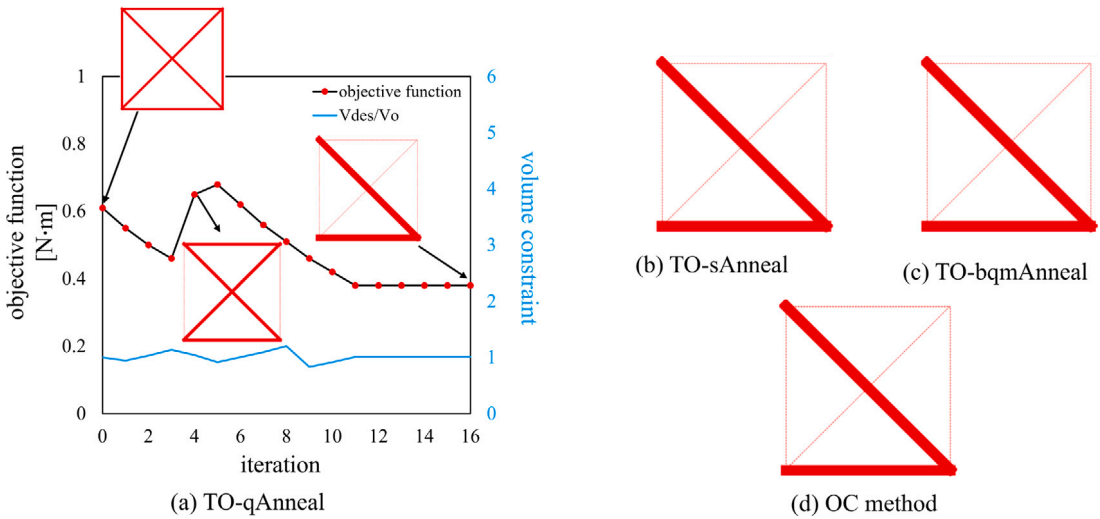

Fig. 6. Optimization results: 6 truss members with different solvers.

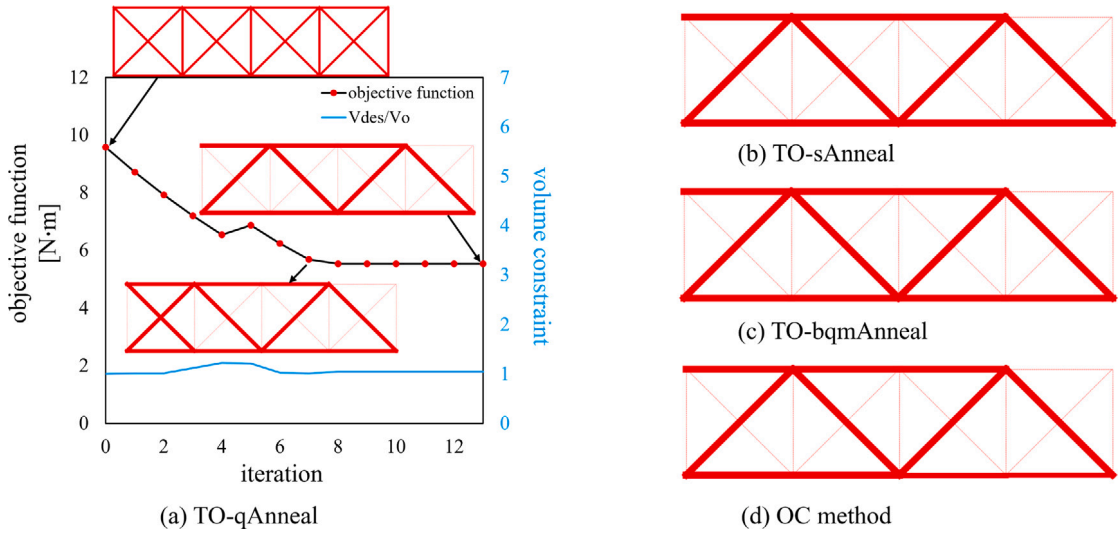

Fig. 7. Optimization results: 21 truss members with different solvers.

Fig. 7 illustrates the optimized results of the initial 21-member truss structure in Fig. 5(b) with the same parameter settings, except that $\rho_o = 0.5$, and the total number of qubits is $N_q = 22$. As can be seen, TO-qAnneal achieves a monotonic convergence in the evolution of the value of the objective function and the volume constraint. Also, the final optimized topologies are the same as those for the TO-sAnneal and TO-bqmAnneal, and agree well with that of the OC method.

For the third target truss structure shown in Fig. 5(c), the initial volume is set at $\rho_o = 0.4$, the total number of qubits is $N_q = 30$, repetition sampling is set to 250 times, and remaining parameters are the same. As can be seen, Fig. 8(a) shows a good convergence to the optimized design obtained from TO-qAnneal, and the final configuration after convergence is the same as those of TO-sAnneal and TO-bqmAnneal shown in Figs. 8(b) and (c) but slightly different from the result of the OC method, as shown in Fig. 8(d). This difference is due to the limited number of candidate solutions for the updater $\alpha$ when using quantum and simulated annealers. Specifically, with the $n_q = 1$ qubit in this example, there are only two possible candidate solutions for $\alpha \in \{0, 1.1\}$ in each iteration. This implies that the update process involves either increasing by 10% or deleting it, iteratively, until the design variable

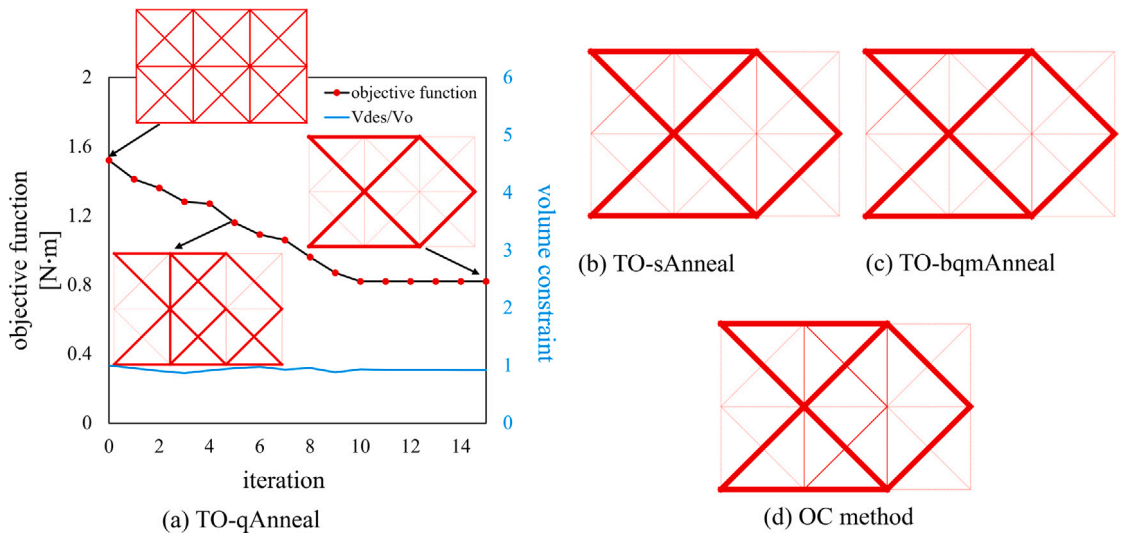

**Fig. 8.** Optimization results: 29 truss members with different solvers.

**Table 2**
Final values of the objective function and time to find the solution in optimization problems of truss structures.

| | | | TO-qAnneal | | | | TO-sAnneal | | | | TO-bqmAnneal | | | OC method |
|---|---|---|---|---|---|---|---|---|---|---|---|---|---|---|---|
| $M$ | $R$ | $I_N$ | $f_{obj}$ | $TFS_{qa}$ (s) | $T_{qpu}$ (s) | $T_{FEM}$ (s) | $f_{obj}$ | $TFS_{sa}$ (s) | $T_{sa}$ (s) | $T_{FEM}$ (s) | $f_{obj}$ | $T_{bqm}$ (s) | $T_{FEM}$ (s) | $f_{obj}$ |
| 6 | 200 | 16 | 0.38 | 0.064 | 0.488 | 0.013 | 0.38 | 0.485 | 160 | 0.015 | 0.38 | 48 | 0.014 | 0.39 |
| 21 | 200 | 13 | 5.53 | 0.052 | 0.422 | 0.023 | 5.53 | 0.556 | 130 | 0.027 | 5.53 | 39 | 0.021 | 5.81 |
| 29 | 250 | 15 | 0.82 | 0.075 | 0.555 | 0.030 | 0.82 | 0.726 | 150 | 0.032 | 0.82 | 45 | 0.029 | 0.82 |

Note: $M$ refers to the total number of truss members and $T_{FEM}$ is the computational time used for structural analysis.

reaches the upper bound and convergence is achieved. As a result, the main members of the optimized structures in Figs. 8(a)–(c) tend to converge to a discrete value, 0 or 1. On the other hand, since the OC method uses solution candidates from a real-valued continuous function, the major members in Fig. 8(d) converge not only to the upper or lower bounds but also to intermediate values as subsidiary members.

Fig. 9 illustrates the convergence of optimal designs obtained from TO-sAnneal and TO-bqmAnneal. As shown, the results are nearly identical and closely align with the QA plotting results. This consistency across computing platforms applies not only to the final optimal outcomes but also throughout the evolution process, especially for TO-sAnneal and TO-bqmAnneal, which are established as ideal platforms for QA applications. These findings confirm the reliability of the developed QA-based framework.

Table 2 compares the final values of the objective function and the time required to find the solution. As can be seen, the objective function values obtained from the annealing methods are slightly different from that of the OC method and tend to be lower, following the same trend reported in [51]. Next, we focus on the computational efficiency of the algorithms, starting with a comparison between QA and SA having the same class of algorithms. As expected, TO-qAnneal takes significantly less time to find a solution in comparison with TO-sAnneal, as confirmed from the comparison between $TFS_{qa}$ and $TFS_{sa}$. However, since quantum hardware is still in its early stages, there is significant overhead time in the QPU, as shown by $T_{qpu}$. Additionally, $TFS_{qa}$ and $TFS_{bqm}$ do not exhibit a straightforward increase with problem size (i.e., the number of qubits), in contrast to the trends observed in classical methods as shown by $TFS_{sa}$. This is one of the advantages of QC in calculation time when scaling problems compared to classical methods. In fact, $TFS_{qa}$ is primarily influenced by the number of repeat samples, $R$, and the number of design iterations, $I_N$. For example, the 21-member truss achieves optimal results with the lowest values of $I_N = 13$ and $R = 200$, resulting in a shorter $TFS_{qa}$ compared to the 6-member truss, which requires $I_N = 16$ and $R = 200$. On the other hand, the 29-member truss, with $I_N = 15$, requires a higher number of repeat samples ($R = 250$), leading to the highest $TFS_{qa}$. It is also worth noting that $TFS_{qa}$ results from running multiple samples on the quantum annealer (e.g., 200 and 250 samples) to mitigate noise and errors from the quantum machine. Therefore, there is a high potential for extreme speedup as quantum devices continue to improve. Indeed, numerous benchmark studies [3,32,51] have highlighted the advantage of QC in terms of time complexity, which refers to how the computation time of an algorithm scales with the size of the input data. Therefore, as the problem size increases, it is expected that TO-qAnneal demonstrates superior performance compared to the classical method. Also, the total limit timeout for SA and the hybrid solver from D-Wave, $T_{sa}$ and $T_{bqm}$, is set by default to be large enough to find possible solutions, so it is not a metric for efficiency comparison but instead provides the actual time usage, including overhead and communication time. While these factors can be minimized through fine-tuning and parametric studies, they will be addressed in future work.

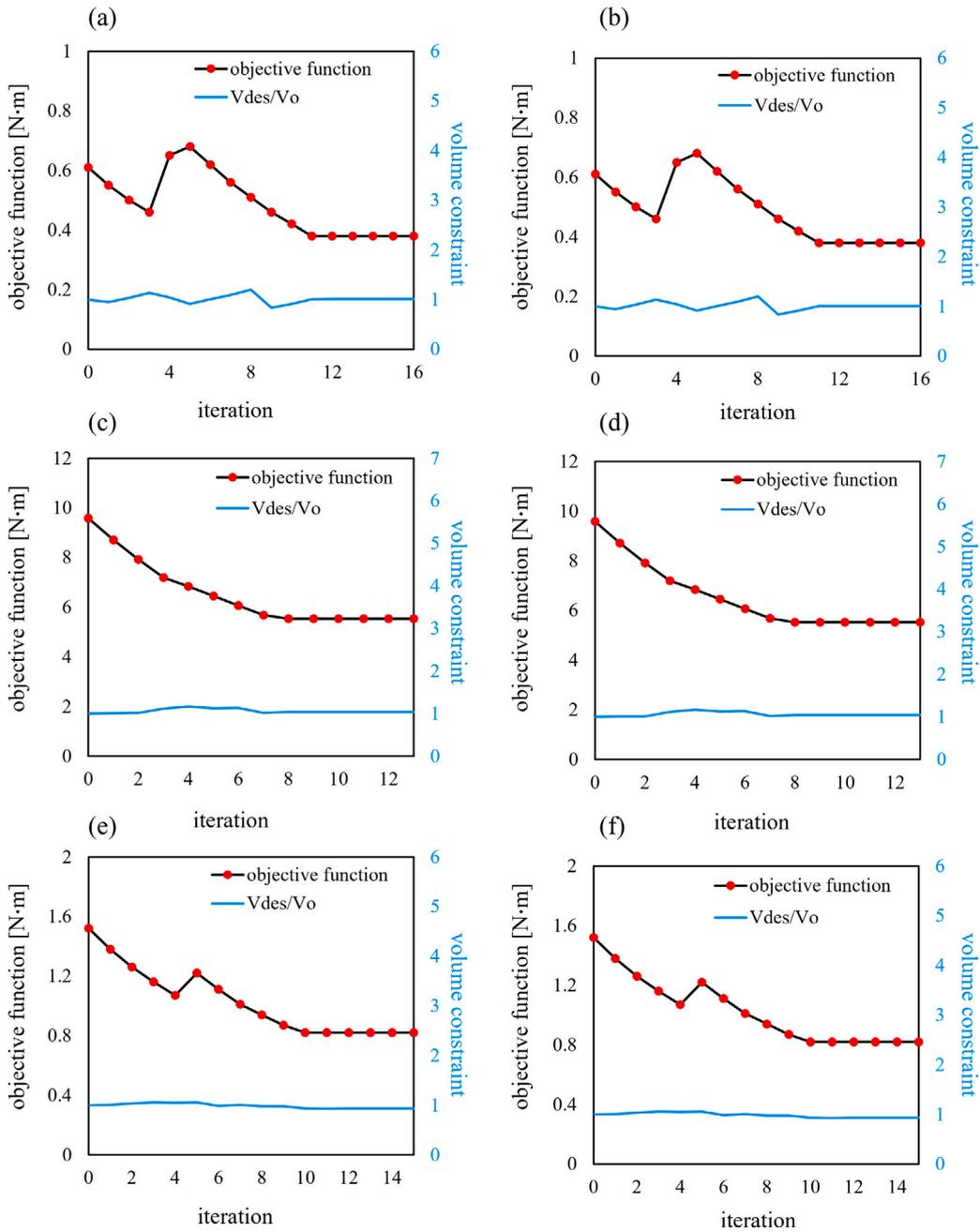

**Fig. 9.** History of the objective function value for truss optimization using `TO-sAnneal` in (a), (c), and (e) and `TO-bqmAnneal` in (b), (d), and (f): (top) 6 members, (middle) 21 members, and (bottom) 29 members.

### 4.2. 2D continuum structure

In this section, we focus our attention on 2D continuum structures made of linearly elastic materials. FE analysis is performed to solve the state variables on a classical computer. Here, the element stiffness matrix in Eq. (8) is calculated as $\mathbf{K}_e^{(i)}(\rho_e) = \int_{\Omega_e} \rho_e^{(i)} \mathbf{B}^T \mathbf{C_0} \mathbf{B} \, dV$, where $\rho_e^{(i)}$ represents the ratio between the current and initial elemental volume, $V_e^{(i)}/V_e^0$, $\mathbf{B}$ is the strain–displacement matrix, $\Omega_e$ is the domain of an element, and $\mathbf{C_0}$ is the elasticity matrix dependent on the material properties. In this study, the isotropic elastic properties are taken as $E = 2 \times 10^{11}$ N/m$^2$ and Poisson's ratio $\nu = 0.3$ under the plane strain condition. Additionally, for this example, $\Theta_s = 0.02$, the target ratio $\bar{V}_{\text{target}}$ is maintained at the initial total volume $V_0$ throughout the optimization process, with $V_e^{(i)}(\rho_e) = \rho_e^{(i)} \cdot V_e^0$.

**Table 3**

Final values of the objective function and time to find the solution in optimization problems of coat-hanging structure.

| | TO-qAnneal | | | | TO-sAnneal | | | | TO-bqmAnneal | | | OC method |
|---|---|---|---|---|---|---|---|---|---|---|---|---|
| | $f_{obj}$ | $TFS_{qa}$ (s) | $T_{qpu}$ (s) | $T_{FEM}$ (s) | $f_{obj}$ | $TFS_{sa}$ (s) | $T_{sa}$ (s) | $T_{FEM}$ (s) | $f_{obj}$ | $T_{bqm}$ (s) | $T_{FEM}$ (s) | $f_{obj}$ |
| coat hanging | 427.7 | 0.06 | 0.570 | 0.306 | 427.7 | 1.011 | 150 | 0.311 | 427.7 | 45 | 0.316 | 465.04 |

Note: $T_{FEM}$ is the computational time used for structural analysis.

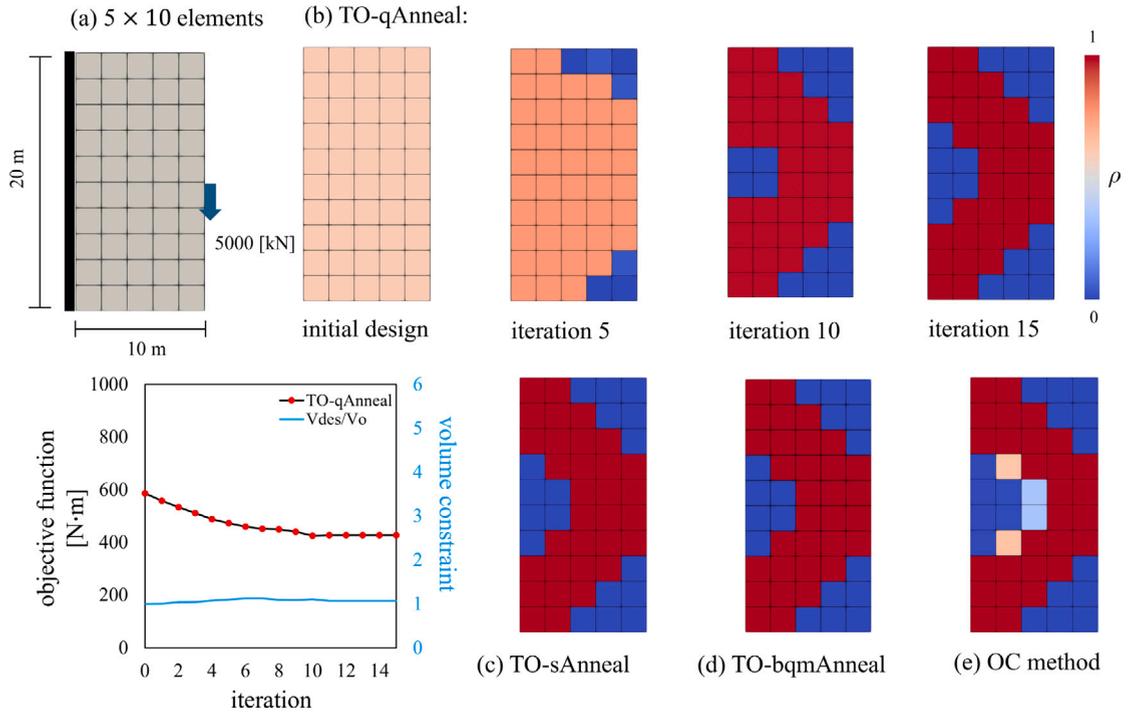

Fig. 10. Optimization results: coat-hanging with different solvers.

First, we target a well-known coat-hanger problem [42], as shown in the leftmost panel of Fig. 10(a), with $\rho_o = 0.6$, $\Theta_e = 1.05$ and $\lambda$ set to 500. Repetition sampling is set to 200 times in each design iteration for TO-qAnneal, while the limit timeout is set by default to 10 and 3 s for TO-sAnneal and TO-bqmAnneal, respectively. The total number of qubits is $N_q = 51$. The optimized topology and its evolution from the proposed design framework run on TO-qAnneal, TO-sAnneal and TO-bqmAnneal are shown in Figs. 10(b)–(d) and Figs. 12(a)–(b), along with the optimized result obtained using the OC method in Fig. 10(e). It can be seen from the figure that the optimization results converge to a similar topology, except that the result from the OC method has some intermediate values or gray elements, which is a characteristic of the density-based approach. While filtering techniques could be adopted to address this issue, it is outside the scope of this study, as we prioritize investigating unperturbed results without introducing additional smoothing techniques. Although the proposed method tends to clearly split the design variable into 0 or 1, the binary encoding process in Eq. (26), coupled with the small number of $n_q$, limits the number of candidate solutions. This limitation can lead to an overestimation of the design volume, but it can be adjusted by modifying the penalty parameter $\lambda$. Meanwhile, the final objective function value and calculation time are shown in Table 3. Once again, the final objective function values of all annealing solvers are equivalent and lower than those obtained by the OC method. As can be seen, the $TFS_{qa}$ is the shortest. Notably, the QPU access time does not significantly increase despite the higher number of qubits compared to the truss optimization example. This evidence supports the advantage of time complexity when exploiting quantum computing, particularly when approaching larger problem sizes. However, as the number of qubits increases, the level of noise also increases, which can lead to nondeterministic solutions. While increasing the number of repetitive samples can help control the noise, it also results in longer computation times. This trade-off poses a limitation on the current quantum hardware to fully investigate the performance of the proposed algorithm.

Next, to demonstrate the applicability of the proposed algorithm to larger problem sizes, we consider a cantilever beam-like solid structure with the dimension as illustrated in the top of Fig. 11. In addition, it is discretized with a mesh resolution of 80 × 40 elements, and the total number of qubits is $N_q = 3201$. Since this problem size exceeds the capacity of the currently available quantum annealer machine from D-wave (for TO-qAnneal), we execute our proposed algorithm using TO-sAnneal and TO-bqmAnneal for this case. The design parameters are set as follows: $\rho_o = 0.6$, $\Theta_e = 1.1$, $\lambda = 8 \times 10^3$, and the time limit is set to 10 s, the remaining parameters are the same as the previous coat-hanging example.

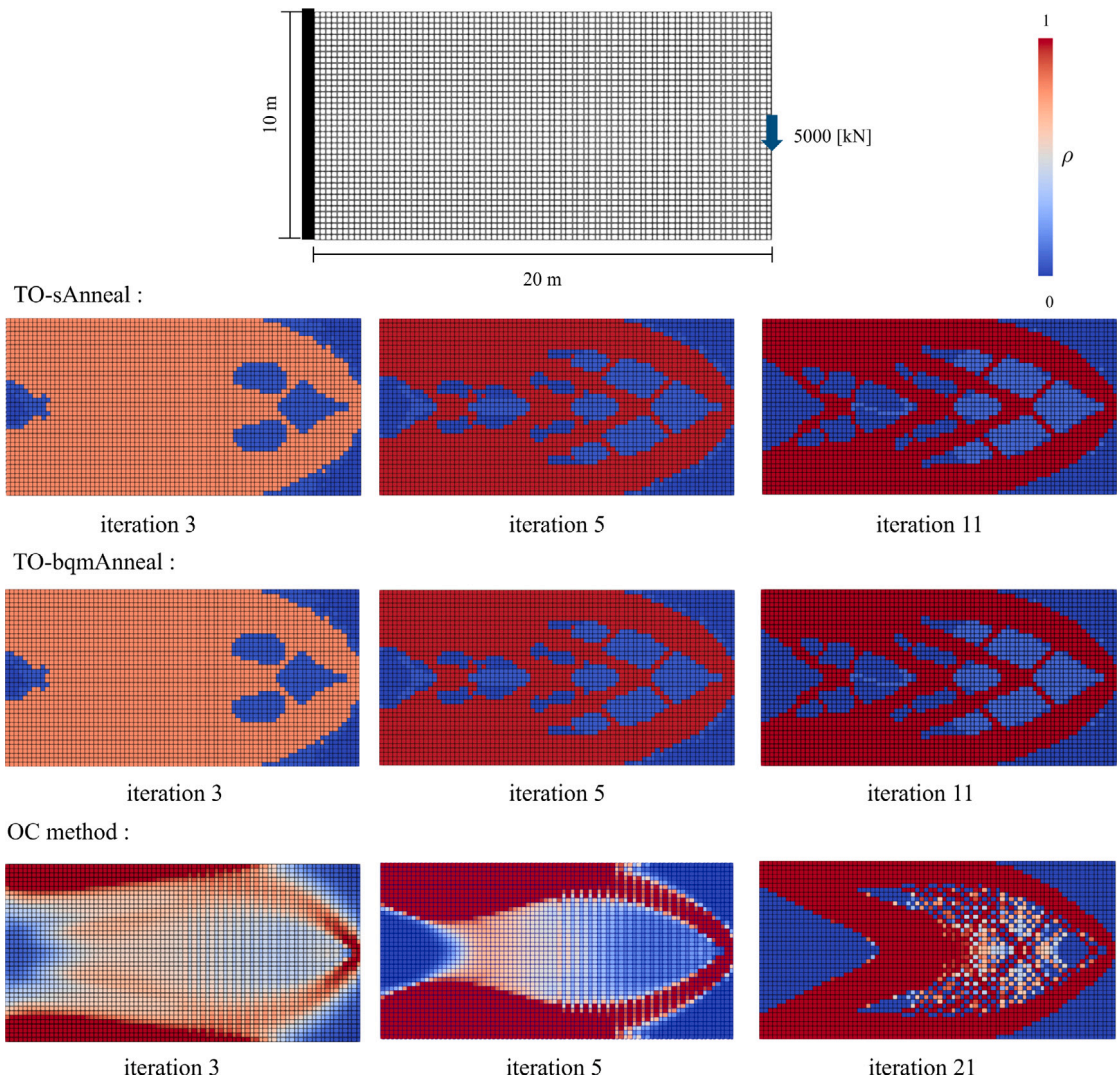

**Fig. 11.** Snapshot of optimization results for 2D cantilever beam-like structure with 80 × 40 elements.

Fig. 11 presents a snapshot of the optimization results, showing that TO-sAnneal and TO-bqmAnneal produce the same results, which implies the reliability of the proposed algorithm. As demonstrated in the previous example, TO-qAnneal, TO-sAnneal and TO-bqmAnneal achieve similar optimal results. Therefore, based on the results of the previous example, we expect that reasonable and comparable results should be obtained when quantum machines are enhanced and ready to handle large-scale problems. In contrast, the OC method faces numerical issues in this case, as some of the elements converge to intermediate values. This is a common numerical instability in topology optimization, which typically requires filtering techniques to resolve. The proposed QA-based framework with a single qubit per element, however, naturally avoids this issue and produces clear designs, relying instead on fine-tuning parameters such as $\lambda$ and $\theta$. This feature may be one of the advantages of our framework in achieving smooth designs.

In addition, Figs. 12(c)–(d) illustrate the histories of the objective function values and design volumes for both solvers, TO-sAnneal and TO-bqmAnneal. The results indicate that the optimization outputs are consistent and converge effectively to the optimal designs. This confirms the reliability and efficiency of the proposed framework for large problem sizes.

### 4.3. 3D continuum structure

In the final example, the proposed algorithm is applied to 3D continuum structures to demonstrate its robustness and performance using TO-sAnneal and TO-bqmAnneal. Additionally, the OC method without filtering is also applied as a reference. However, it should be noted that the results may differ due to the complexity of the design problem and the differences between optimization algorithm approaches. Cubic and L-shaped design domains are targeted and discretized with $20 \times 20 \times 20$ and $40 \times 40 \times 5$ elements, respectively, as shown in Fig. 13 and Fig. 14. As in the previous examples, FE analysis is conducted on a classical computer to solve the state variables, where the stiffness matrix formulation is similar to the 2D problem but accounts for much larger degrees of freedom than 2D cases. However, the volume constraint is still effectively enforced and can be adjusted by modifying the penalty value. The material properties are defined with the Young's modulus of $E = 2 \times 10^{11} \, \text{N/m}^2$ and a Poisson's ratio of $\nu = 0.3$. Additionally, the parameters are set as $\Theta_e = 1.1$ and $\Theta_s = 0.02$.

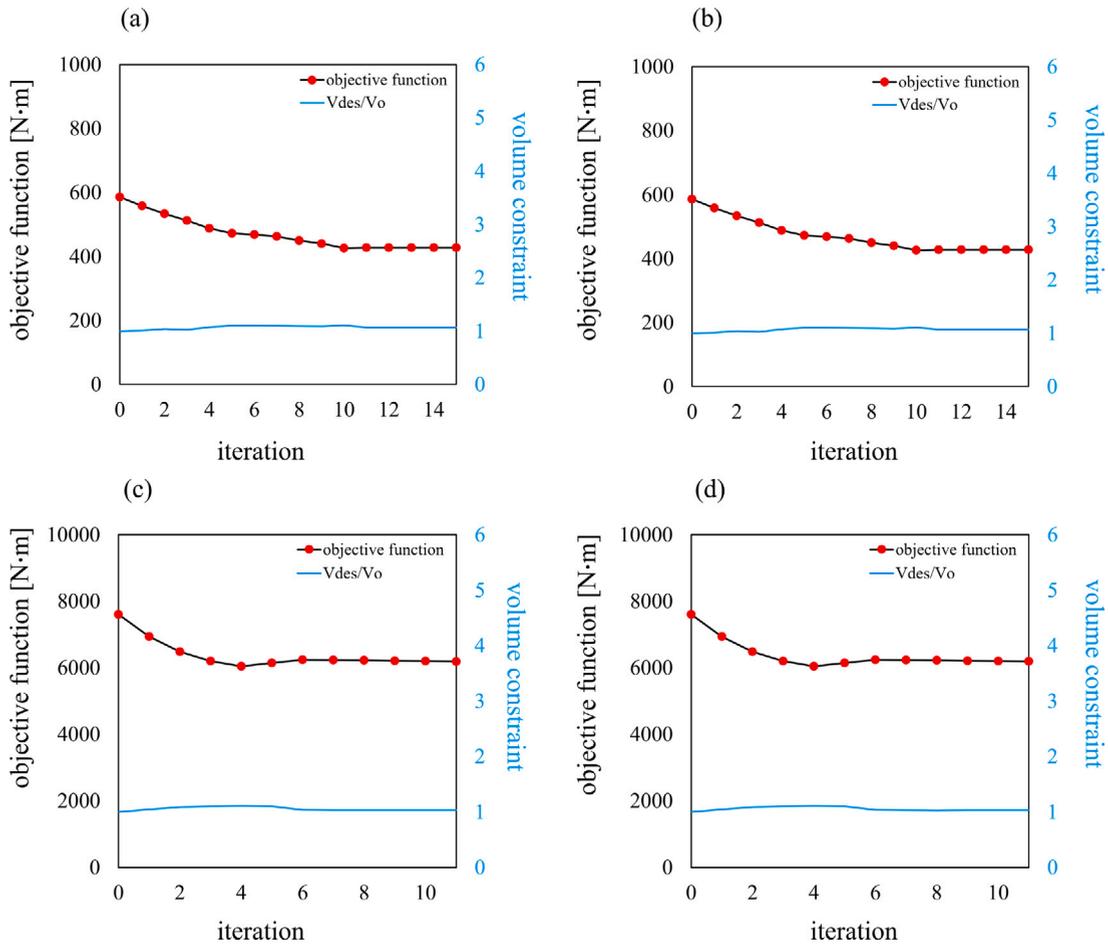

**Fig. 12.** History of the objective function values for (top row) a coat-hanging and (bottom row) a 2D cantilever beam-like structure, with `TO-sAnneal` in (a) and (c), and `TO-bqmAnneal` in (b) and (d).

First, we consider a cube subjected to tension at the center of the top surface at $z = 10 \, \text{m}$, with a fixed boundary condition at the bottom surface $z = 0 \, \text{m}$. The initial volume, maximum allowable value and penalty parameter are set to $\rho_o = 0.25$, and $\lambda = 1000$, respectively. The limit timeout is set to the default of 10 s, and the total number of qubits used is $N_q = 8001$. The optimized topology at design variable $\rho > 0.5$ is displayed in Figs. 13, along with the plotted history of the objective function value and volume constraint presented in the top row of Fig. 15. As can be seen from these figures, the optimization converges well for both the optimal topology and objective function value within 20 iterations. Once again, the final optimized topologies from `TO-sAnneal` and `TO-bqmAnneal` are identical, including the evolution of objective function values. Additionally, the volume constraint is strictly maintained throughout the optimization process, even though it has been formulated as a QUBO model, which is inherently unconstrained. As can be seen, the optimized design from the OC method shows a similar tendency to those from `TO-sAnneal` and `TO-bqmAnneal`. Notably, the final objective function value from the OC method is 810.0 [N m], which is slightly lower than the 812.3 [N m] obtained from `TO-sAnneal` and `TO-bqmAnneal`. This difference is due to the final design volume of the OC method (250 m³) being larger than that of `TO-sAnneal` and `TO-bqmAnneal`. The design volume can be adjusted by fine-tuning the penalty constant value $\lambda$; however, a detailed parametric study will be addressed in future work.

Next, consider an L-shaped structure subjected to a downward distributed force applied on the region defined by $x = 80 \, \text{m}$, $z = 40 \, \text{m}$, and $y = 0$ to $10 \, \text{m}$, with a fixed boundary condition at the plane $z = 80 \, \text{m}$. The parameters are set as follows: $\rho_o = 0.5$, and $\lambda = 10^5$. The time limit is set to the default of 10 s, and the total number of qubits used is $N_q = 6001$. The optimized topology for design variables $\rho > 0.5$ is shown in Figs. 14. As shown, the results indicate that both TO-sAnneal and TO-bqmAnneal yield similar optimal topology designs, with only slight differences in the objective function evolution. Both methods strictly maintain the volume constraints, as demonstrated in the bottom row of Fig. 15. Nevertheless, the optimal result obtained differs from that of the OC method which can occur with complex boundary conditions (e.g., Ye et al. [51]), such as stress concentrations in this case. Specifically, the final objective function value and volume for the OC method are 35106.6 [N m] and 24000 m³ (forming only the exterior surface), respectively. In contrast, the final objective function values for TO-sAnneal and TO-bqmAnneal are 36032.6 and 35966.6 [N m], with design volumes of 23861.4 and 23846 m³, respectively.

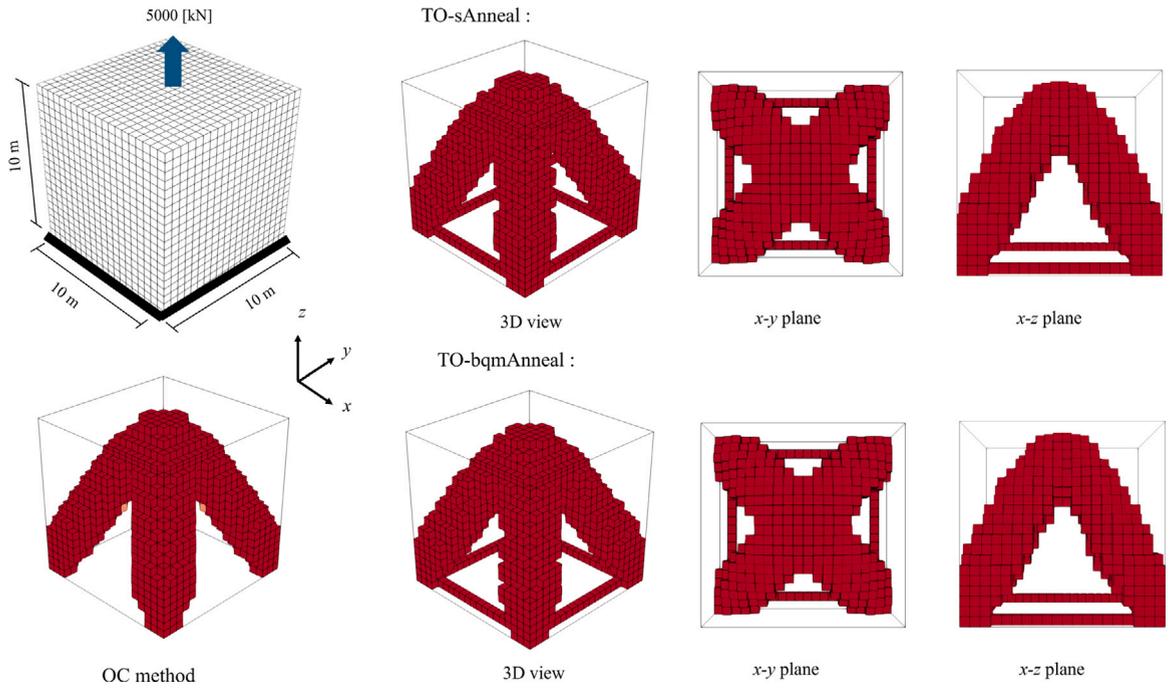

**Fig. 13.** Optimized topology at $\rho > 0.5$: A 3D cube under tension discretized with $20 \times 20 \times 20$ elements.

It should be noted that a single qubit per element, $n_q = 1$, is used to represent the unknown variables (i.e., $\alpha(q_e)$ and $\bar{S}(q_s)$). Consequently, the set of candidate solutions is limited, especially for the slack variable $\bar{S}$. However, the volume constraint is still effectively enforced as shown in Fig. 15 and can be adjusted by modifying the penalty value $\lambda$. This example not only highlights the versatility of the proposed algorithm but also underscores its scalability when applied to larger problem sizes. Although the quantum annealer (TO-qAnneal) is currently limited by hardware constraints, the consistent performance with TO-sAnneal and TO-bqmAnneal suggests that the proposed quantum-based algorithms hold significant potential to solve even more complex and large-scale optimization problems as quantum hardware continues to advance.

It can be argued that this is the first study to address a 3D continuum structure with such a large number of elements and qubits (e.g., $N_q = 8001$), demonstrating the promising scalability and capability of the proposed framework. As the technology matures, quantum annealers are expected to overcome current limitations, enabling the application of these algorithms to higher-resolution problems and further demonstrating the robustness and efficiency of the proposed approach.

## 5. Conclusions

This study presents a novel design update framework for topology optimization with quantum annealing, employing elemental updaters concept along with mapping to the corresponding design variables. The framework is particularly advantageous due to its simplicity and computational efficiency, facilitating fast convergence to optimal designs. A QUBO formulation was derived for the compliance minimization problem, constrained by an inequality volume condition. A series expansion was used to encode the conversion between real and binary values for the design variables, i.e. updaters, enabling the output variables in the QUBO model to be updated as their product.

The performance of the developed quantum annealing-based design framework was demonstrated on discrete trusses as well as on a simple 2D continuum coat-hanger problem, using a quantum annealer and validated against results from a hybrid classical-quantum annealer, a GPU-based Ising machine (simulated annealer), and the optimality criteria method. Additionally, to demonstrate scalability, the proposed quantum annealing-based framework was applied to a 2D cantilever beam and 3D continuum models, such as a cube and an L-shaped structure, using both a hybrid classical-quantum annealer and a simulated annealer. The optimization results confirmed that the proposed quantum annealing-based design framework reliably converged to the optimal design. Remarkably, even with a limited number of binary variables, or equivalently, a small number of qubits, the proposed quantum annealing-based approach effectively converged to optimal designs. Our numerical results indicate that, although computation on quantum annealing computers remains challenging in this emerging research field, the computational speed of quantum annealing in finding a solution is faster than that of simulated annealing in the GPU-based Ising machine, confirming the advantages provided by quantum effects. However, quantum computers have yet to surpass classical computers due to overheads, such as communication time between external systems and the internal QPU. With the anticipated rapid development of quantum devices, the integration of the quantum annealing-based framework with topology optimization holds promising potential for the near future.

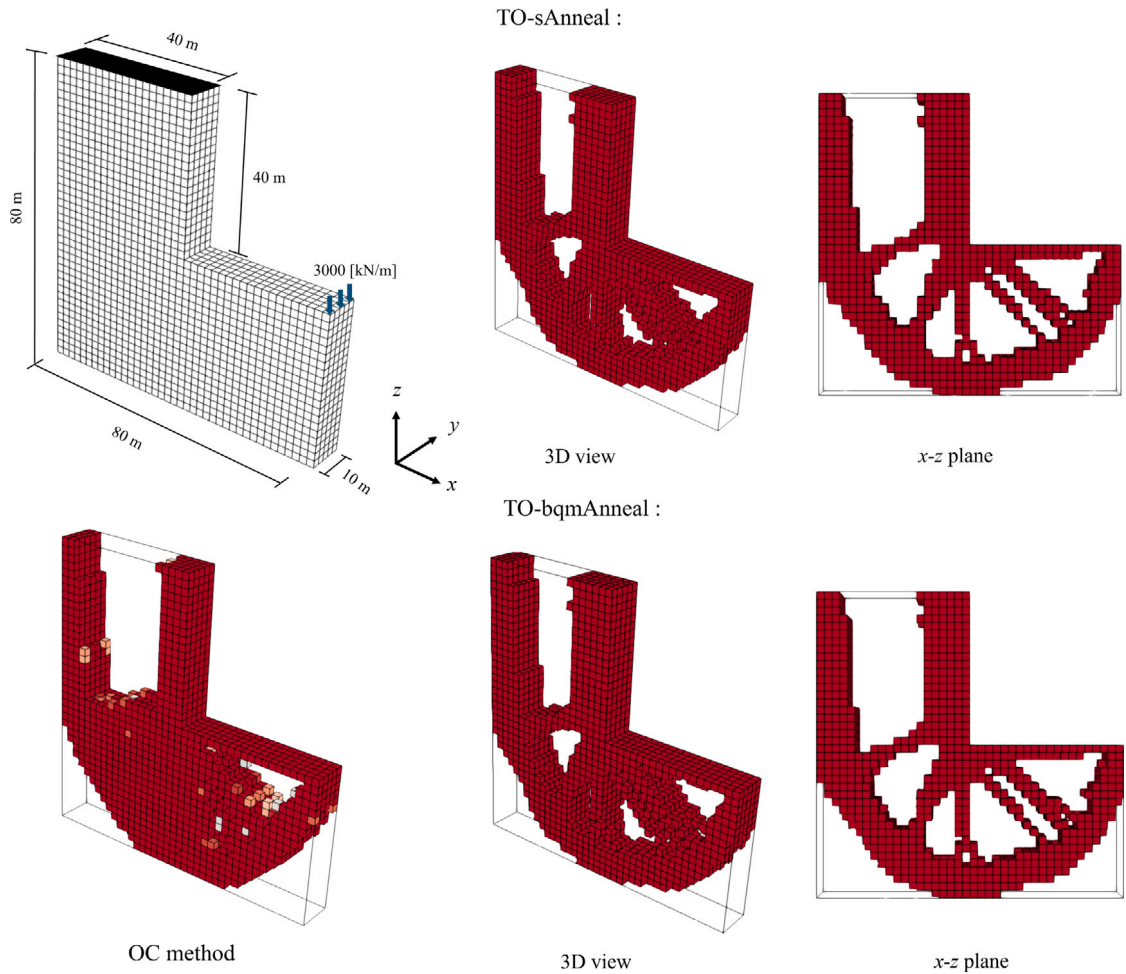

**Fig. 14.** Optimized topology at $\rho > 0.5$: A L-shaped under uniform downward force discretized with $40 \times 5 \times 40$ elements.

It should be noted that the parameters within the proposed framework require fine-tuning for each specific problem, particularly the penalty parameter for imposing the volume constraint in the QUBO model. Therefore, further development is needed to automate the process of finding optimal parameter values. Furthermore, exploring alternative functional forms and the influence of the encoding process could further increase the efficiency of updating the design variable. In future work, the developed framework should be extended to conduct structural analysis using quantum annealing, in order to improve computational efficiency, as this is the dominant factor affecting computational time.

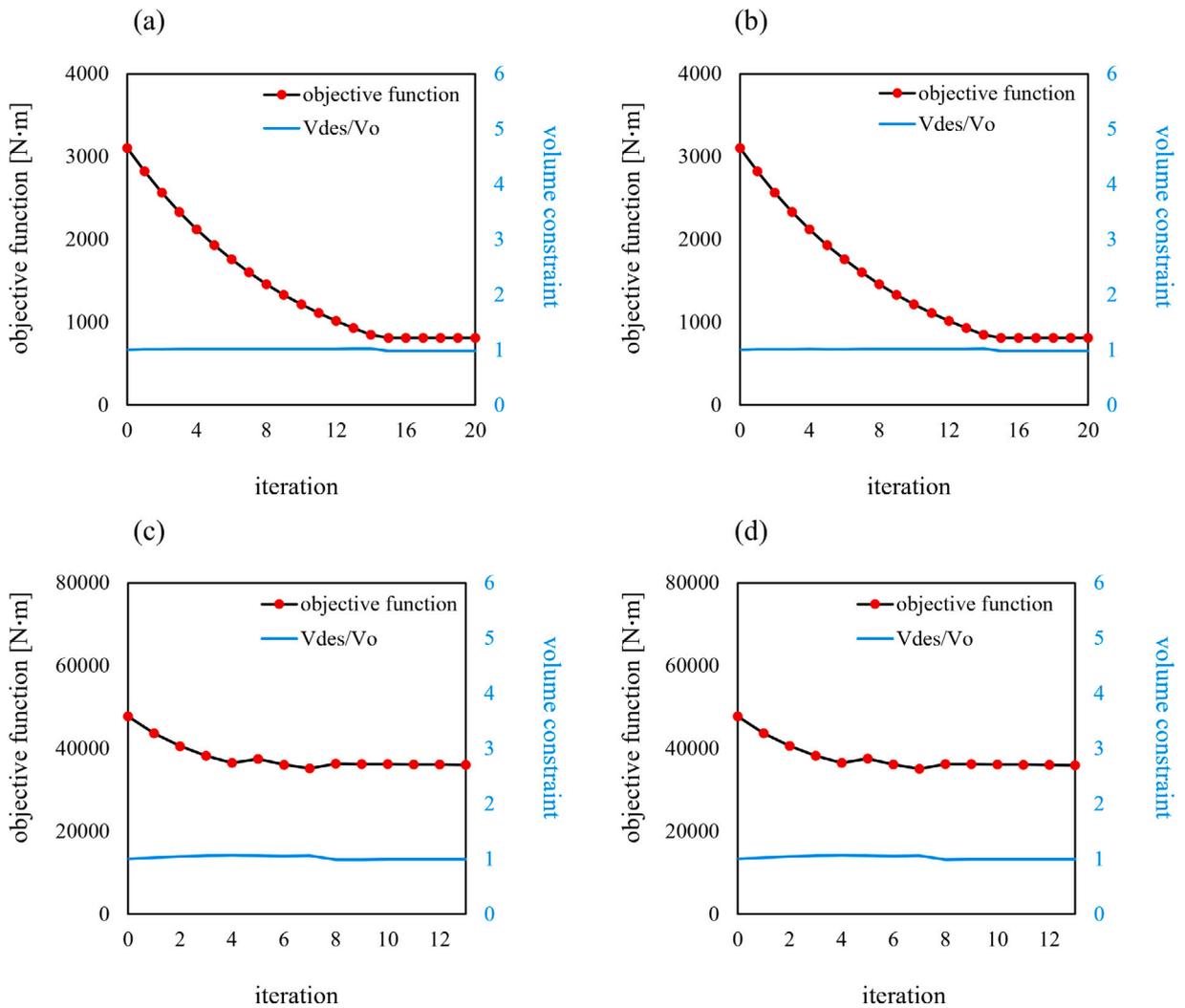

**Fig. 15.** History of the objective function values for (top row) a cube and (bottom row) an L-shaped structure, with `TO-sAnneal` in (a) and (c), and `TO-bqmAnneal` in (b) and (d).

**CRediT authorship contribution statement**

**Naruethep Sukulthanasorn:** Writing – original draft, Visualization, Validation, Software, Resources, Methodology, Investigation, Formal analysis, Data curation, Conceptualization. **Junsen Xiao:** Writing – review & editing, Visualization, Methodology, Investigation. **Koya Wagatsuma:** Writing – review & editing, Software, Methodology, Data curation. **Reika Nomura:** Writing – review & editing, Visualization. **Shuji Moriguchi:** Writing – review & editing, Supervision. **Kenjiro Terada:** Writing – review & editing, Supervision, Project administration, Methodology, Investigation, Funding acquisition, Conceptualization.

**Declaration of competing interest**

The authors declare that they have no known competing financial interests or personal relationships that could have appeared to influence the work reported in this paper.

**Data availability**

Data will be made available on request.